\documentclass[journal]{IEEEtran}
\ifCLASSINFOpdf

\else

\fi

\usepackage[T1]{fontenc}
\usepackage{color, colortbl}
\usepackage{babel}
\usepackage{verbatim}
\usepackage{textcomp}
\usepackage{amsmath}
\usepackage{amsthm}
\usepackage{amssymb}
\usepackage{amstext}
\usepackage{graphicx}
\usepackage{tabulary}
\usepackage{caption}
\usepackage{subcaption}
\usepackage{epstopdf}
\usepackage{cite}
\usepackage{balance}
\usepackage{multirow}
\usepackage{xcolor}
\usepackage{soul}
\usepackage[ruled,vlined]{algorithm2e}
\addto\captionsenglish{}
\usepackage{bm}
\usepackage{bbm}
\usepackage{soul}
\usepackage{booktabs}

\usepackage{slashbox}

\newtheorem{Proposition}{\bf Proposition}
\newtheorem{Lemma}{\bf Lemma}

\newtheorem{Algorithm}{\bf Algorithm}

\def\RSS{\mathtt{RSS}}

\def\SINR{\mathtt{SINR}}

\newcommand{\green}[1]{{\textcolor[rgb]{0,0.5,0}{#1}}}

\DeclareMathOperator*{\argmax}{arg\,max}

\allowdisplaybreaks

\begin{document}

\bstctlcite{IEEEexample:BSTcontrol}

\title{Mathematical Cell Deployment Optimization for Capacity and Coverage of Ground and UAV Users}

\author{Saeed Karimi-Bidhendi, 
        Giovanni Geraci, and
        Hamid Jafarkhani
\thanks{Saeed Karimi-Bidhendi and Hamid Jafarkhani are with the Center for Pervasive Communications \& Computing, University of California, Irvine, Irvine CA, 92697 USA (e-mail: \{skarimib, hamidj\}$@$uci.edu).
{Their work was supported in part by the NSF Award CNS-2229467.}
}
\thanks{Giovanni~Geraci is with Telef\'{o}nica Scientific Research and Universitat Pompeu Fabra (UPF), Barcelona, Spain (e-mail: giovanni.geraci$@$upf.edu). 
His work was supported by the HORIZON-SESAR-2023-DES-ER-02 project ANTENNAE (101167288), by the Spanish State Research Agency through grants CNS2023-145384, PID2021-123999OB-I00, and CEX2021-001195-M, and by the UPF-Fractus Chair on Tech Transfer and 6G.
} 
}

\maketitle

\begin{abstract}
We present a general mathematical framework for optimizing cell deployment and antenna configuration in wireless networks, inspired by quantization theory. Unlike traditional methods, our framework supports networks with deterministically located nodes, enabling modeling and optimization under controlled deployment scenarios.
We demonstrate our framework through two applications: joint fine-tuning of antenna parameters across base stations (BSs) to optimize network coverage, capacity, and load balancing, and the strategic deployment of new BSs, including the optimization of their locations and antenna settings. These optimizations are conducted for a heterogeneous 3D user population, comprising ground users (GUEs) and uncrewed aerial vehicles (UAVs) along aerial corridors.
Our case studies highlight the framework's versatility in optimizing performance metrics such as the coverage-capacity trade-off and capacity per region. Our results confirm that optimizing the placement and orientation of additional BSs consistently outperforms approaches focused solely on antenna adjustments, regardless of GUE distribution. Furthermore, joint optimization for both GUEs and UAVs significantly enhances UAV service without severely affecting GUE performance.
\end{abstract}

\begin{IEEEkeywords}
Deployment optimization, cellular networks, quantization theory, UAV corridors, drones, aerial highways.
\end{IEEEkeywords}

\IEEEpeerreviewmaketitle

\section{Introduction}\label{Introduction}

\subsection{Motivation and Related Work}

The coverage and capacity of cellular networks are significantly influenced by the deployment sites of cells and the configuration of base station (BS) antennas. Adjustments in parameters such as the downtilt angle are crucial for optimizing signal strength and minimizing interference. This process, known as cell shaping, is complex because of the inter-dependencies among the settings across multiple cells. Suitable cell deployment and properly tuned antenna parameters enhance signal reception in critical cell areas while reducing interference with neighboring cells.

Optimizing BS deployment sites and antenna settings is inherently challenging. The settings across cells are coupled by interference, making the optimization problem nonconvex and NP-hard \cite{TekNovAko2024}. Additionally, there are conflicting objectives to consider: maximizing coverage probability, which typically involves directing energy towards cell edges, and maximizing capacity, which favors high signal-to-interference-plus-noise ratio (SINR) for cell-center users. Balancing these objectives for a heterogeneous population of users, including aerial users flying along corridors, adds another layer of complexity \cite{CheJaaYan2020,BerLopPio2023}. Indeed, optimizing for aerial coverage requires directing some of the radiated energy upwards, conflicting with the needs of legacy ground users, who benefit from downtilted cells \cite{GerGarAza2022,wu20205g}. 

Existing approaches to optimizing cell deployment rely on various techniques, each with its own limitations. In the Third Generation Partnership Project (3GPP), global optimization methods based on stochastic system simulations are used \cite{3GPP36777,3GPP38901}. These simulations typically apply to small, homogeneous hexagonal layouts where exhaustive search techniques can determine fixed values, such as uniform downtilt angles across all cells. However, these methods do not generalize to real-world networks with diverse and complex configurations. In actual networks, site-specific radio frequency planning tools are employed, relying heavily on trial-and-error methods and field measurements. Not only are these approaches time-consuming, but also they fail to achieve scalable and near-optimal solutions. These shortcomings are aggravated in more challenging scenarios with diverse user populations, including uncrewed aerial vehicles (UAVs) \cite{GerGarGal2018,GarGerLop2019,DanGarGer2020}. 

More advanced techniques have been explored to address these challenges, such as reinforcement learning (RL) \cite{BalAnd2019,10609329,
VanIakHak2021,DanShaKle2017,BouFarFor2021} and Bayesian optimization (BO) \cite{DreDauQia2021,BenGerLop2023}, showing potential in BS deployment optimization. RL, with its ability to adapt to dynamic environments, seems promising but requires substantial data to achieve accuracy and has slow convergence rates, leading to extensive computations and prolonged simulations \cite{TekNovAko2024}. Additionally, RL lacks safe exploration, as its random exploration strategies can lead to suboptimal antenna parameter configurations that degrade system performance. 
Conversely, BO offers faster convergence and safer exploration but suffers from high computational complexity, making it suitable only for low-dimensional problems \cite{ShaSweWan2015,frazier2018tutorial,MagValHoy2021}. Despite these advancements, a general framework for optimizing cellular networks to maximize key performance indicators (KPIs) effectively is still missing. This paper aims to fill this gap by proposing a new mathematical approach, inspired by quantization theory, to cell deployment optimization.


\subsection{Contribution and Summary of Results}

In this paper, we develop a mathematical framework for optimal cell deployment and antenna configuration leveraging quantization theory~\cite{gray1998quantization}, already proven successful in addressing problems that involve the geographical deployment of agents. Unlike stochastic geometry---commonly used for the statistical analysis of wireless networks with random topologies \cite{haenggi2009stochastic, win2009mathematical, elsawy2013stochastic }, including UAV networks  \cite{azari2019cellular,AzaGerGar2020}---quantization theory is particularly effective in analyzing structured networks with a finite number of deterministically located nodes.  This mathematical framework enables accurate modeling and optimization of network performance under controlled deployments. It has been successfully applied to the optimal deployment of antenna arrays \cite{koyuncu2018performance}, access point placement for the optimal throughput \cite{gopal2022modified, gopal2023vector}, and power optimization in wireless sensor networks \cite{GuoJaf2016, cortes2005spatially, guo2018source, ingle2011energy, guo2019movement, cortes2004coverage, karimi2020energy, tang2019three, karimi2021energy, wang2006movement, karimi2023outage}. Applications of quantization theory have also been extended to UAV communications, including trajectory optimization and deployment of UAVs \cite{koyuncu2018deployment,9086619}, optimal UAV placement for rate maximization \cite{shabanighazikelayeh2020optimal}, and the deployment of UAVs as power efficient relay nodes \cite{koyuncu2018power}.

In this paper, we build upon our previous contributions \cite{KarGerJaf2023} to develop a general mathematical framework for cell deployment optimization. This framework enables the fine-tuning of antenna parameters for each BS within a given deployment to achieve optimal coverage, capacity, or any trade-off thereof. Additionally, it allows to optimize cell partitioning accounting for load balancing across cells, ensuring a more efficient distribution of network traffic. Moreover, the framework allows for the optimization of additional BS placements, both in terms of cell site locations and antenna parameters.

Specifically, we determine the necessary conditions and design iterative algorithms to optimize (i) the vertical antenna tilts and transmit power at each existing BS in a cellular network, and (ii) the location, vertical tilt, horizontal bearing, and transmit power of each newly deployed BS, to provide the best quality of service to a given geographical distribution of users. Our analysis accommodates multiple KPIs, supporting a flexible 3D user distribution for each KPI considered. This flexibility allows, for instance, the prioritization of a specific KPI (e.g., sum-log-capacity) on the ground and another (e.g., coverage probability) along 3D aerial corridors, or any desired trade-off between the two. To the best of our knowledge, this is the first work to rigorously and tractably address these optimization goals while accounting for realistic network deployment, antenna radiation patterns, and propagation channel models.

{
To demonstrate the effectiveness of our mathematical framework, we present two case studies, where we optimize: (i) the coverage-capacity trade-off and (ii) the capacity per region. We maximize these metrics for a heterogeneous user population, comprising ground users (GUEs), distributed either uniformly or following a Gaussian mixture, and UAVs along 3D aerial corridors. 
Key insights from these case studies are as follows:
\begin{itemize}
\item 
Optimizing the locations and bearings of additional BSs consistently outperforms solely optimizing vertical antenna tilts and transmission powers, regardless of the GUE distribution.
\item 
Jointly optimizing the network for both GUEs and UAVs yields significant service improvements for UAVs with minimal performance impact on GUEs, compared to optimizing for GUEs alone.
\end{itemize}
}

\section{General Framework and Example Applications}\label{System_Model}

In this section, we present our general mathematical framework based on quantization theory and illustrate it with two example applications. The first example focuses on optimizing the configuration of an existing set of cellular BSs. The second example extends this by optimizing both the deployment and configuration of new BSs. We then introduce the system model and the KPIs we aim to maximize for both applications.

\subsection{General Mathematical Framework}\label{General_Framework}

We consider a terrestrial cellular deployment consisting of $N$ BSs that provide service to network users located in the 3D space. The $N$ cellular BSs are characterized by certain parameters $\bm{\alpha}_1, \cdots, \bm{\alpha}_K$, where $\bm{\alpha}_k \in \mathbb{R}^{N_k}$ is an object of optimization for each $k \in \{1, \cdots, N\}$ and $N_k$ is the dimension of vector $\bm{\alpha}_k$. 

Let $\lambda(\bm{q})$ be a probability density function that represents the distribution of users over a 3D target region $Q$. For the simplicity of presentation, we assume that $\lambda(\bm{q})$ is known and independent of time; however, our proposed framework is equally applicable to dynamic setting where $\lambda(\bm{q})$ varies with time. Each user is associated with one BS; thus, the target region $Q$ is partitioned into $N$ disjoint subregions $\bm{V} = (V_1, \cdots, V_N)$ such that users within $V_n$ are associated with BS $n$.

In its simplest formulation, our problem is to maximize a certain performance function $\mathcal{P}$, given by:
\begin{equation}\label{objective-function}
    \mathcal{P}(\bm{V}, \bm{\alpha}_1, \ldots, \bm{\alpha}_K) = \sum_{n=1}^{N} \int_{V_n} \gamma(\bm{q}; \bm{\alpha}_1, \ldots, \bm{\alpha}_K) \lambda(\bm{q}) d\bm{q}
\end{equation}
over the cell partitioning $\bm{V}$ and the parameters $\{\bm{\alpha}_1, \cdots, \bm{\alpha}_K\}$, for a given KPI $\gamma(\cdot)$ and distribution $\lambda(\bm{q})$. Not only does the optimal choice of each parameter depend on the value of the other, but also this optimization problem is NP-hard. Our approach is to design alternating optimization algorithms that iterate between updating $\bm{V}$ and $\{\bm{\alpha}_1, \cdots, \bm{\alpha}_K\}$.

Although the probability density function $\lambda(\bm{q})$ is continuous and stochastic, the same approach works with discrete or finite user locations, both for stochastic and known locations. Such a discrete distribution of users will replace the integrals in \eqref{objective-function} with summations over the corresponding values. The  derivations and algorithms in the rest of this manuscript will work for the discrete case with the modified performance function.

In quantization theory, variations of the Lloyd algorithm \cite{lloyd1982least, gray1998quantization} have been used to solve similar optimization problems and provide analytical insights. When designed using data instead of density function models, these algorithms can be categorized as unsupervised learning methods similar to the K-means algorithm. Inspired by quantization theory, we aim to maximize $\mathcal{P}(\bm{V}, \bm{\alpha}_1, \ldots, \bm{\alpha}_K)$ in (\ref{objective-function}) through alternating optimization by: (i) finding the optimal cell partitioning $\bm{V}$ given a set of parameters $\{\bm{\alpha}_1, \cdots, \bm{\alpha}_K\}$; and (ii) finding the optimal parameters $\{\bm{\alpha}_1, \cdots, \bm{\alpha}_K\}$ for a given cell partitioning $\bm{V}$. 
The solution of the first task is a generalized Voronoi tessellation \cite{boots2009spatial, du1999centroidal}. For the second task, our approach is to apply gradient ascent to find the optimal parameters $\{\bm{\alpha}_1, \cdots, \bm{\alpha}_K\}$ for a given cell partitioning $\bm{V}$. Gradient ascent is a first-order optimization algorithm that iteratively refines the estimate of locally optimal $\{\bm{\alpha}_1, \cdots, \bm{\alpha}_K\}$ by taking small steps in the direction of the gradient.
In the following, we provide two practical applications of our mathematical framework for radio access network optimization.

\subsection{Application~\#1: Base Station Antenna Tilt Optimization and Power Allocation for Ground and UAV Users}\label{Tilt_Optimization}

In the first application, we apply our mathematical framework to jointly optimize the vertical antenna tilts and the transmission powers of all BSs to enhance the capacity and coverage on a target 3D region $Q$. The system model considered for the first application follows the 3GPP specifications in \cite{3GPP38901,3GPP36777} and is detailed as follows.

\begin{figure}
\centering
\includegraphics[width=\columnwidth]{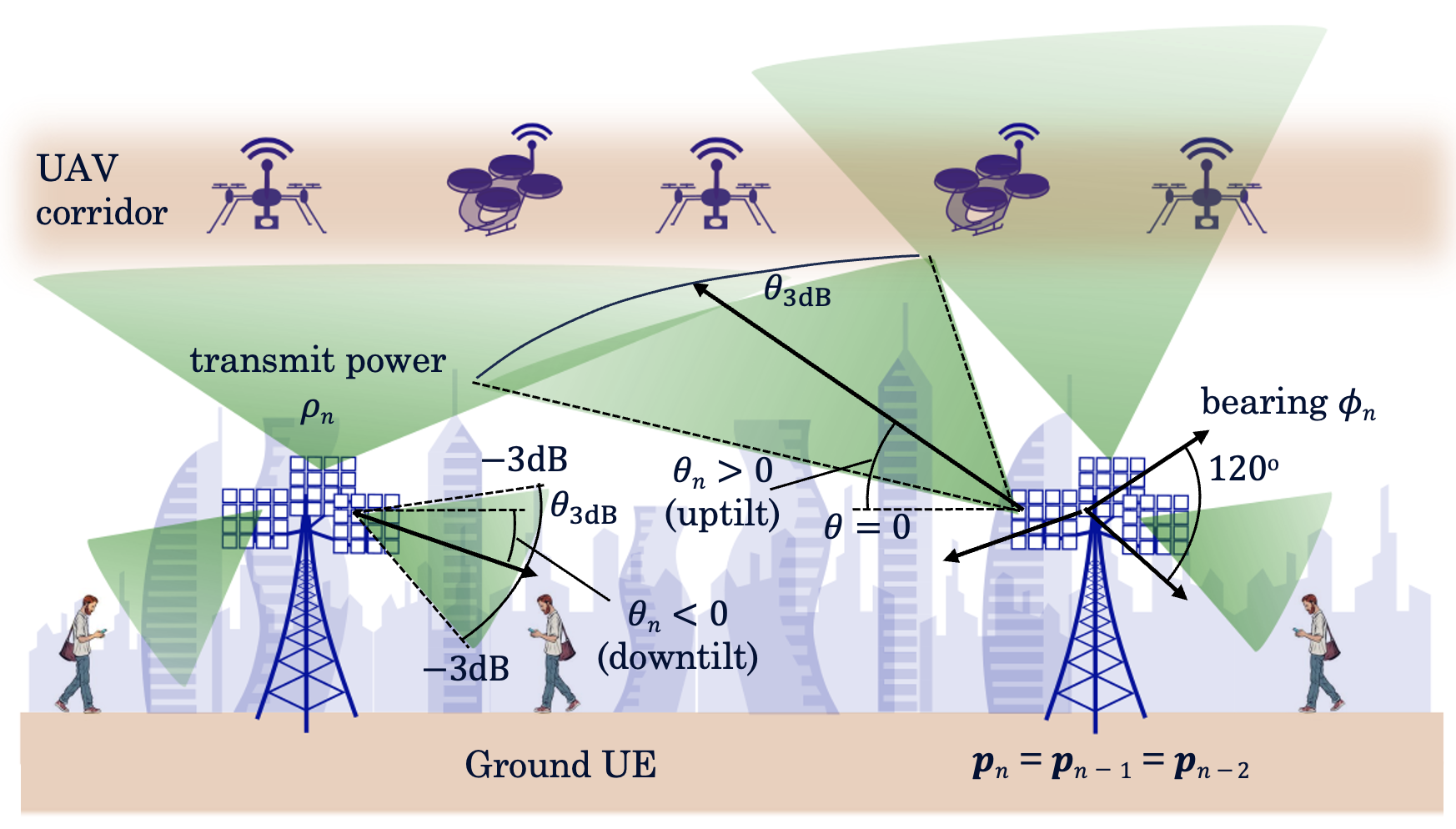}
\caption{Uptilted ($\theta_n>0$) and downtilted ($\theta_n<0$) BSs serving GUEs and UAV corridors, with $\theta_n$, $\theta_{\textrm{3dB}}$, $\phi_n$, and $\rho_n$ denoting vertical tilt, vertical HPBW, horizontal bearing, and transmit power, respectively.  Also, $p_{n-2} = p_{n-1} = p_n$ denote the position of three co-located sectorized BSs.}
\label{fig:illustration}
\end{figure}

\subsubsection*{Ground cellular network}
The location of BS $n$ is denoted by  $\bm{p}_n \in \mathbb{R}^2$, for each $n \in \{1, \cdots, N\}$. Let $\bm{\Theta} = (\theta_1, \cdots, \theta_N)$, where $\theta_n \in [-90^\circ, +90^\circ]$ is the vertical antenna tilt of BS $n$ that can be electrically adjusted by a mobile operator, with positive and negative angles denoting uptilts and downtilts, respectively. Let $\bm{\rho} = (\rho_1, \cdots, \rho_N)$, where $\rho_n$ is the transmission power of BS $n$, measured in dBm, which is also adjustable by a mobile operator with a maximum value of $\rho_{\max}$. We denote the antenna horizontal bearing (in the azimuth direction) of BS $n$ by $\phi_n \in [-180^\circ, +180^\circ]$, which is assumed to be fixed upon deployment. 

\subsubsection*{Performance function}
For Application~\#1, we assume that the BS location $\bm{p}_n$ and the azimuth orientation $\phi_n$ are fixed for all $n \in \{1, \cdots, N\}$. Thus, the performance function $\mathcal{P}$ in (\ref{objective-function}) becomes:
\begin{equation}\label{general_objective_no_deployment}
\mathcal{P}^{(1)}_{\gamma}(\bm{V}, \bm{\Theta}, \bm{\rho}) = \sum_{n=1}^{N} \int_{V_n} \gamma^{(n)}(\bm{q}; \bm{\Theta}, \bm{\rho}) \lambda(\bm{q})d\bm{q},
\end{equation}
and for a given KPI $\gamma(\cdot)$, the goal is to optimize this performance function over the cell partitioning $\bm{V}=(V_1,\cdots,V_N)$, BS vertical antenna tilts $\bm{\Theta} = (\theta_1, \cdots, \theta_N)$, and BS transmission powers $\bm{\rho} = (\rho_1, \cdots, \rho_N)$. 
Fig.~\ref{fig:illustration} illustrates a representative scenario and the main corresponding system parameters.

\subsection{Application~\#2: Base Station Deployment Optimization for Ground and UAV Users}\label{Deployment_Optimization}

In this second application, in addition to optimizing the configuration of the existing set of BSs, we seek to apply our mathematical framework to optimally deploy and configure new BSs. To this end, we assume that there are $N - M$ BSs with fixed locations and bearings, denoted as $\underline{\bm{P}} = (\bm{p}_1, \cdots, \bm{p}_{N-M})$ and $\underline{\bm{\Phi}} = (\phi_1, \cdots, \phi_{N-M})$, respectively. We seek to deploy $M$ additional BSs, whose locations and bearings are parameters to be optimized, denoted as $\overline{\bm{P}} = (\bm{p}_{N-M+1}, \cdots, \bm{p}_{N})$ and $\overline{\bm{\Phi}} = (\phi_{N-M+1}, \cdots, \phi_{N})$, respectively. Hence, the location of all BSs and their bearings are given by $\bm{P} = (\bm{p}_1, \cdots, \bm{p}_{N})$ and $\mathbf{\Phi} = (\phi_1, \cdots, \phi_{N})$, respectively. The vertical antenna tilts associated with all BSs are given by $\mathbf{\Theta} = (\theta_1, \cdots, \theta_{N})$ and their transmission powers by $\bm{\rho} = (\rho_1, \cdots, \rho_{N})$. Finally, the cell partitioning is given by $\mathbf{V} = (V_1, \cdots, V_{N})$.

\subsubsection*{Sectorized cellular deployment}

While our framework can accommodate any customized BS deployment, we make the common assumption of sectorized cells in our applications. More precisely, each deployment site consists of three BSs that share the same position and have equally spaced bearings. 
Let $\widehat{\bm{p}}_m$ and $\widehat{\phi}_m$ denote the location of the $m$-th deployment site and its reference bearing, respectively. The locations and bearings for the three co-located cells $n = 3m$, $n-1$, and $n-2$ are therefore given by: 
\begin{align}\label{entangled_BS_locations}
    \widehat{\bm{p}}_m &= \bm{p}_{n-2} = \bm{p}_{n-1} = \bm{p}_{n}, \\
    \widehat{\phi}_m &= \phi_{n-2} = \phi_{n-1} - 120^\circ = \phi_{n} - 240^\circ. \label{entangled_BS_orientations}
\end{align}
We define 
$\widehat{\bm{P}} = \big(\widehat{\bm{p}}_1, \cdots, \widehat{\bm{p}}_{\frac{N}{3}} \big)$ and $\widehat{\bm{\Phi}} = \big(\widehat{\phi}_1, \cdots, \widehat{\phi}_{\frac{N}{3}}  \big)$ for compact representations.

\subsubsection*{Performance function}
For Application \#2 and a given KPI $\gamma(\cdot)$, the performance function $\mathcal{P}$ in (\ref{objective-function}) becomes:
\begin{equation}\label{general_objective_with_deployment}
\mathcal{P}^{(2)}_{\gamma}\big(\bm{V}, \bm{\Theta}, \bm{\rho}, \overline{\widehat{\bm{P}}}, \overline{\widehat{\bm{\Phi}}}\big) \!=\! \sum_{n=1}^{N} \int_{V_n}\! \gamma^{(n)}\big(\bm{q}; \bm{\Theta}, \bm{\rho}, \overline{\widehat{\bm{P}}}, \overline{\widehat{\bm{\Phi}}}\big) \lambda(\bm{q})d\bm{q},
\end{equation}
where $\overline{\widehat{\bm{P}}} = \Big(\widehat{\bm{p}}_{\frac{N - M}{3} + 1}, \cdots, \widehat{\bm{p}}_{\frac{N}{3}} \Big)$ and $\overline{\widehat{\bm{\Phi}}} = \Big(\widehat{\phi}_{\frac{N - M}{3} + 1}, \cdots, \widehat{\phi}_{\frac{N}{3}} \Big)$.
The goal is thus to maximize the above performance function over the vertical antenna tilts $\bm{\Theta}$ and transmission powers $\bm{\rho}$ of all BSs, the cell partitioning $\bm{V}$, as well as the locations $\overline{\widehat{\bm{P}}}$
and bearings $\overline{\widehat{\bm{\Phi}}}$ of the additional BSs to be deployed. 

In the remainder of the paper, we maximize $\mathcal{P}^{(1)}_{\gamma}$ and $\mathcal{P}^{(2)}_{\gamma}$ for two KPIs, $\gamma_1(\cdot)$ and $\gamma_2(\cdot)$, defined as follows.

\subsection{Channel Model and KPIs}
\label{section:KPIs}

We now introduce our propagation channel model and performance metrics such as the SINR and the achievable rate. From those, we then define the two KPIs 
used in our applications.

\subsubsection*{Antenna gain} 
BSs use directional antennas with vertical and horizontal half-power beamwidths of $\theta_{\textrm{3dB}}$ and $\phi_{\textrm{3dB}}$, respectively. Let $A_{\max}$ be the maximum antenna gain at the boresight and denote the vertical and horizontal antenna gains in dB at user location $\bm{q} \in Q$ by $A_{n,\bm{q}}^{\mathrm{V}}$ and $A_{n,\bm{q}}^{\mathrm{H}}$, respectively. Directional antenna gains are given by \cite{3GPP38901}:
\begin{equation}\label{directional-antenna-gains}
    A_{n,\bm{q}}^{\mathrm{V}} =
    - \frac{12}{\theta^2_{\text{3dB}}} \left[ \theta_{n,\bm{q}} - \theta_n \right]^2, \quad \!\!
    A_{n,\bm{q}}^{\mathrm{H}} =
    - \frac{12}{\phi^2_{\text{3dB}}} \left[ \phi_{n,\bm{q}} - \phi_n \right]^2, 
\end{equation}
where $\theta_{n,\bm{q}}$ and $\phi_{n,\bm{q}}$ are the elevation angle and the azimuth angle between BS $n$ and the user at $\bm{q}$, respectively, in degrees. These angles can be calculated as:
\begin{align}\label{theta-phi-nq}
    \theta_{n,\bm{q}} &= \tan^{-1}\!\left( \frac{{q_{\mathrm{z}}} - {p_{n, \mathrm{z}}} }{{\sqrt{(q_\mathrm{x} - p_{n, \mathrm{x}})^2 + (q_\mathrm{y} - p_{n, \mathrm{y}})^2 }}  }  \right), \\
    \phi_{n,\bm{q}} &= \!
\begin{cases}
\tan^{-1}\!\left(\frac{q_{\mathrm{y}}-p_{n,\mathrm{y}}}{q_{\mathrm{x}}-p_{n,\mathrm{x}}}\right) \!+\! 180^{\circ}\!\times\! 2c &  \!\!\!\text{if $q_{\mathrm{x}} \!-\! p_{n,\mathrm{x}}>0$}\\
\tan^{-1}\!\left(\frac{q_{\mathrm{y}} - p_{n,\mathrm{y}}}{q_{\mathrm{x}}-p_{n,\mathrm{x}}}\right) \!+\! 180^{\circ}\!\times\! (2c + 1) &  \!\!\!\text{if $q_{\mathrm{x}} \!-\! p_{n,\mathrm{x}}<0$}
\end{cases}\nonumber
\end{align}
where subscripts $\cdot_{\mathrm{x}}$ and $\cdot_{\mathrm{y}}$ denote the horizontal and vertical Cartesian coordinates of a point, respectively, {and $\cdot_{\mathrm{z}}$ denotes its height.}  The integer $c$ is selected such that $-180^{\circ} \leq \phi_{n,\bm{q}} - \phi_n \leq +180^{\circ}$. Thus, the total antenna gain of BS $n$ in dB is given by $A_{n, \bm{q}} = A_{\max} + A_{n,\bm{q}}^{\mathrm{V}} +A_{n,\bm{q}}^{\mathrm{H}}$.

\subsubsection*{Pathloss} 
The pathloss $L_{n,\bm{q}}$ between BS $n$ and the user location $\bm{q}$ is a function of their 3D distance and given by:
\begin{equation} \label{eqn:Pathloss}
L_{n,\bm{q}} = a_{\bm{q}} + b_{\bm{q}} \log_{10} \big( \| \bm{q} - \bm{p}_n \|  \big),
\end{equation}
where $a_{\bm{q}}$ depends on the carrier frequency while $b_{\bm{q}}$ depends on the line-of-sight condition and the pathloss exponent, and in turn on the BS deployment scenario and the height of the user at $\bm{q}$. In our case studies, we utilize practical values for the constants $a_{\bm{q}}$ and $b_{\bm{q}}$ that are adopted from the 3GPP \cite{3GPP36777,3GPP38901}.

\subsubsection*{Received signal strength}

In our applications, we aim at optimizing long-term cell planning decisions. To this end, we focus on large-scale fading, since small-scale fading affects instantaneous signal quality and its effect is typically mitigated via channel coding. 
The wideband received signal strength (RSS) from BS $n$, measured in dBm, provided at the user location $\bm{q}$ is given by:
\begin{multline}\label{RSS-dBm}
\mathtt{RSS_{dBm}^{(n)}}(\bm{q})= \rho_n + A_{n,\bm{q}} - L_{n,\bm{q}}
 = \rho_n + A_{\textrm{max}} \\ - \frac{12}{\theta^2_{\text{3dB}}} \left[ \theta_{n,\bm{q}} - \theta_n \right]^2 - \frac{12}{\phi^2_{\text{3dB}}} \left[ \phi_{n,\bm{q}} - \phi_n \right]^2 
  \\ - a_{\bm{q}}  - b_{\bm{q}} \log_{10} \big( {\| \bm{q} - \bm{p}_n \| } \big). 
\end{multline}

\subsubsection*{Wideband SINR}

Using the definition of $\mathtt{RSS_{dBm}^{(n)}}$ in (\ref{RSS-dBm}), we define the wideband signal-to-interference-plus-noise ratio (SINR) in dB as:
\begin{equation}\label{eqn:SINR_dB}
\mathtt{SINR^{(n)}_{dB}} (\bm{q}) = 10 \log_{10} \frac{10^{\frac{1}{10}\mathtt{RSS^{(n)}_{dBm}} (\bm{q})}}{\sum_{j\neq n}^{} 10^{\frac{1}{10}\mathtt{RSS^{(j)}_{dBm}} (\bm{q})} + \sigma^2},
\end{equation}
where $\sigma^2$ denotes the thermal noise variance in linear units.

\subsubsection*{Achievable rate}

{The spectral efficiency at the user location $\bm{q}$, expressed in bps/Hz, is given by:
\begin{equation}
R(\bm{q}) = \log_2\big(1 + {\sf SINR_{lin}^{(n)}}(\bm{q}) \big),
\label{eqn:spectral_efficiency}
\end{equation}
where $n$ is the index of BS to which the user at $\bm{q}$ is associated. For simplicity, we refer to (\ref{eqn:spectral_efficiency}) as the achievable rate in the rest of the paper. The actual data rate is determined by multiplying the spectral efficiency by the system bandwidth.}

\subsubsection*{KPI \#1 --- Coverage-capacity trade-off}

our first KPI consists of the two following terms.

\subsubsection{Sum-log-rate} widely used in the literature to strike a balance between sum-rate maximization and fairness across users, and given by:
\begin{equation}\label{sum_log_rate_metric}
    \gamma'^{(n)}(\bm{q}) = \log_2\Big( \log_2\big(1 + {\sf SINR_{lin}^{(n)}}(\bm{q}) \big) \Big).
\end{equation}
Maximizing the sum-log-rate, as opposed to the sum-rate, prevents degenerate solutions where the KPI is dominated by a small fraction of users. 

\subsubsection{Coverage} commonly used to ensure a desired percentage of users receive a minimum SINR $T$ required to maintain connectivity, given by:
\begin{equation}\label{exact_outage_definition}
    \gamma''^{(n)}(\bm{q}) = \mathbbm{1} \big({\sf SINR_{dB}^{(n)}}(\bm{q}) \geq T \big),
\end{equation}
and approximated via a differentiable function as follows:
\begin{equation}\label{approximate_outage_definition}
    \widetilde{\gamma}''^{(n)}(\bm{q}) = \frac{1}{1 + e^{-\kappa \big[{\sf SINR_{dB}^{(n)}}(\bm{q}) - T \big]}},
\end{equation}
where the parameter $\kappa$ controls the steepness of the sigmoid approximation to the indicator function.

\noindent The first KPI is then defined as: 
\begin{equation}\label{second_KPI}
    \gamma_1^{(n)}(\bm{q}) = \beta \gamma'^{(n)}(\bm{q}) + (1 - \beta) \widetilde{\gamma}''^{(n)}(\bm{q}),
\end{equation}
where the parameter $\beta$ creates a trade-off between the sum-log-rate and the coverage.

\subsubsection*{KPI \#2 --- Capacity per region}

Our second KPI is the rate per region across the network. 
%
%
In its most general form, this KPI is given by:
\begin{equation}\label{first_KPI_biased_capacity_per_region}
    \gamma_2^{(n)}(\bm{q}) = \frac{\log_2 \big(1 + {\sf SINR_{lin}^{(n)}}(\bm{q}) \big)}{o_n + \int_{V_n} \lambda(\Tilde{\bm{q}})d\Tilde{\bm{q}}},
\end{equation}
where $\bm{o} = (o_1, \cdots, o_N)$ is an offset hyperparameter to prevent degenerate configurations where tiny cells dominate the network resources.


The two KPIs $\gamma_1(\cdot)$ and $\gamma_2(\cdot)$ are studied in detail in Sections  \ref{sum_log_rate_section} and \ref{capacity_per_region_section}, respectively, for both applications introduced in Sections~\ref{Tilt_Optimization} and \ref{Deployment_Optimization}.


\section{Optimizing Cellular Networks for Coverage-capacity Trade-offs}\label{sum_log_rate_section}

In this section, we study the two applications introduced in Sections \ref{Tilt_Optimization} and \ref{Deployment_Optimization} for the first KPI, $\gamma_1^{(n)}$, given in (\ref{second_KPI}).


\subsection{Application \#1: Base Station Antenna Tilt Optimization and Power Allocation for Coverage-Capacity Trade-off}\label{sum_log_rate_application_1_section}

For the coverage-capacity trade-off, KPI $\gamma_1^{(n)}$ in (\ref{second_KPI}), the maximization of (\ref{general_objective_no_deployment}) becomes:
\begin{align}\label{SumLogRate_App1_performance_function}
&\max_{\substack{\bm{V}, \bm{\Theta}, \bm{\rho} \\ \bm{\rho}\in \bm{\Lambda} } } \mathcal{P}^{(1)}_{\gamma_1}(\bm{V}, \bm{\Theta}, \bm{\rho}) 
\nonumber \\ 
=&\max_{\substack{\bm{V}, \bm{\Theta}, \bm{\rho} \\ \bm{\rho}\in \bm{\Lambda} } } 
\Bigg\{
\beta \sum_{n=1}^{N} \int_{V_n} \!\log_2\Big( \log_2\big(1 + {\sf SINR_{lin}^{(n)}}(\bm{q}) \big) \Big) \lambda(\bm{q})d\bm{q} \nonumber\\
&+ (1 - \beta) \sum_{n=1}^{N} \int_{V_n} \frac{1}{1 + e^{-\kappa \big[{\sf SINR_{dB}^{(n)}}(\bm{q}) - T \big]}} \lambda(\bm{q})d\bm{q} \Bigg\},
\end{align}
where $\bm{\Lambda} = (-\infty, \rho_{\max}]^N$ is the feasible region for the vector of transmission powers $\bm{\rho}$.

The goal of (\ref{SumLogRate_App1_performance_function}) is to reach an optimal coverage-capacity trade-off with respect to the cell partitioning $\bm{V}$, the BS vertical antenna tilts $\bm{\Theta}$, and the BS transmission power $\bm{\rho}$. 
These parameters are interdependent in the sense that the optimal value for each parameter depends on the value of others. In quantization theory, variants of the Lloyd algorithm \cite{lloyd1982least, gray1998quantization} have been applied to similar NP-hard problems. Inspired by the success of these approaches, we devise an alternating optimization algorithm that iterates between optimizing $\bm{V}$, $\bm{\Theta}$, and $\bm{\rho}$. More specifically,  we seek to: (i) find the optimal cell partitioning $\bm{V}$ for given $\bm{\Theta}$ and $\bm{\rho}$; (ii) find the optimal BS vertical antenna tilts $\bm{\Theta}$ for given $\bm{V}$ and $\bm{\rho}$; and (iii) find the optimal BS transmission power $\bm{\rho}$ for given $\bm{V}$ and $\bm{\Theta}$.

\subsubsection*{Optimal cell partitioning}

We begin by describing the optimization process for the cell partitioning $\bm{V}$.

\begin{Proposition}\label{optimal_cell_partitioning_sum_log_rate}
The optimal cell partitioning $\bm{V}^*$ that maximizes the performance function in (\ref{SumLogRate_App1_performance_function})  is given by:
\begin{align}\label{optimal_cell_partitioning_sum_log_rate_eq}
    V^*_n = \{\bm{q} \in Q \mid {\sf RSS_{dBm}^{(n)}}(\bm{q}) \geq {\sf RSS_{dBm}^{(k)}}(\bm{q}), \textrm{ }\textrm{ }  \forall k \neq n \},
\end{align}
for each $n \in \{1, \cdots, N \}$, where ties are broken arbitrarily.
\end{Proposition}

\textit{Proof.} We first demonstrate the following lemma.
\begin{Lemma}\label{ascending_func_of_sinr}
Let $f: \mathbb{R}^+ \rightarrow \mathbb{R}$ be a continuous and increasing function. For a given set of $\left(\bm{V}, \bm{\Theta}, \bm{\rho}, \bm{P},  \bm{\Phi}\right)$, let:
\begin{align}
    F\left(\bm{V}, \bm{\Theta}, \bm{\rho}, \bm{P},  \bm{\Phi}\right) = \sum_{n = 1}^{N} \int_{V_n} f\big({\sf SINR_{dB}^{(n)}}(\bm{q}) \big)\lambda(\bm{q})d\bm{q}.
\end{align}
Then, the optimal cell partitioning that maximizes $F$ for a given set of $\left(\bm{\Theta}, \bm{\rho}, \bm{P}, \bm{\Phi}\right)$ is given by (\ref{optimal_cell_partitioning_sum_log_rate_eq}).
\end{Lemma}

\textit{Proof. }Let $\bm{V}^* = \argmax_{\bm{V}}F\left(\bm{V}, \bm{\Theta}, \bm{\rho}, \bm{P},  \bm{\Phi}\right)$ denote the cell partitioning that maximizes $F$. Then, we have:
\begin{align}\label{optimal_cells_for_F}
    V^*_n &= \{\bm{q} \mid  f\big({\sf SINR_{dB}^{(n)}}(\bm{q}) \big) \geq f\big({\sf SINR_{dB}^{(k)}}(\bm{q}) \big), \quad \forall k \neq n  \} \nonumber\\& 
    = \{\bm{q} \mid  {\sf SINR_{dB}^{(n)}}(\bm{q}) \geq {\sf SINR_{dB}^{(k)}}(\bm{q}), \quad \forall k \neq n  \} \nonumber \\&
    = \{\bm{q} \mid  {\sf RSS_{dBm}^{(n)}}(\bm{q}) \geq {\sf RSS_{dBm}^{(k)}}(\bm{q}), \quad \forall k \neq n  \}.
\end{align}
The last equality in (\ref{optimal_cells_for_F}), which follows from the proof of Proposition 4 in \cite{KarGerJaf2023}, has the same expression as in (\ref{optimal_cell_partitioning_sum_log_rate_eq}) and concludes the proof of Lemma \ref{ascending_func_of_sinr}. \qed

The proof of Proposition \ref{optimal_cell_partitioning_sum_log_rate} then follows from Lemma \ref{ascending_func_of_sinr} and the fact that both functions $f\big({\sf SINR_{dB}^{(n)}}(\bm{q}) \big) = \log\big(\log(1 + {\sf SINR_{lin}^{(n)}}(\bm{q})) \big)$ and $f\big({\sf SINR_{dB}^{(n)}}(\bm{q}) \big) = \frac{1}{1 + e^{-\kappa \big[{\sf SINR_{dB}^{(n)}}(\bm{q}) - T \big]}}$ are continuous and increasing. $\hfill \blacksquare$


\subsubsection*{Optimal vertical antenna tilts}

Next, we apply gradient ascent to optimize the BS vertical antenna tilts $\bm{\Theta}$ for a given $\bm{V}$ and $\bm{\rho}$. 
%
%
The following proposition provides the expression for the partial derivatives of $\mathcal{P}^{(1)}_{\gamma_1}(\bm{V}, \bm{\Theta}, \bm{\rho})$ w.r.t. the BS vertical antenna tilts.

\begin{Proposition}\label{theta_gradient_sumlograte_outage_function}
    The partial derivative of the performance function $\mathcal{P}^{(1)}_{\gamma_1}(\bm{V}, \bm{\Theta}, \bm{\rho})$ in (\ref{SumLogRate_App1_performance_function}) w.r.t. the BS vertical antenna tilt $\theta_n$ is given by:
\begin{align}\label{gamma_1_1_partial_derivative_wrt_theta}
    &\frac{\partial}{\partial \theta_n} \mathcal{P}^{(1)}_{\gamma_1} = \sum_{m=1}^{N} \int_{V_m} \Bigg [ \frac{\beta \cdot \log_2(e) \cdot \log_2(e) \cdot \ln{10} \cdot 0.1}{\log_2\big(1 + {\sf SINR_{lin}^{(m)}}(\bm{q})\big)}\times \nonumber\\& \frac{{\sf SINR_{lin}^{(m)}}(\bm{q})}{1 + {\sf SINR_{lin}^{(m)}}(\bm{q})}  + \kappa (1 - \beta) \cdot \sigma\Big(\kappa \big[{\sf SINR_{dB}^{(m)}}(\bm{q}) - T \big] \Big) \nonumber\\& \times\! \bigg[1 - \sigma\Big(\kappa \big[{\sf SINR_{dB}^{(m)}}(\bm{q}) - T \big] \Big)\bigg] \Bigg ] \!\cdot\! \frac{\partial}{\partial \theta_n} {\sf SINR_{dB}^{(m)}}(\bm{q}) \lambda(\bm{q})d\bm{q},
\end{align}
where
\begin{equation}\label{partial_sinr_theta_derivatives}
    \frac{\partial {\sf SINR_{dB}^{(m)}}(\bm{q})}{\partial \theta_n} = 
\begin{cases}
    \frac{24}{\theta^2_{\sf 3dB}} (\theta_{n,\bm{q}} - \theta_n), \textrm{ } \textrm{ } \qquad\qquad\qquad\! \text{if } m = n &\\
    - \frac{\frac{24}{\theta^2_{\sf 3dB}}\cdot (\theta_{n,\bm{q}} - \theta_n)\cdot {\sf RSS_{lin}^{(n)}}(\bm{q}) \cdot {\sf SINR_{lin}^{(m)}}(\bm{q}) }{{\sf RSS_{lin}^{(m)}}(\bm{q}) },      \!    \textrm{ }  \quad   \text{o.w.} &
\end{cases}  
\end{equation}
and $\sigma(x) = \frac{1}{1 + e^{-x}}$ is the sigmoid function.
\end{Proposition}

\noindent The proof of Proposition \ref{theta_gradient_sumlograte_outage_function} is provided in Appendix \ref{Appendix_B}.

\subsubsection*{Optimal transmission powers}

Finally, we describe the optimization process for BS transmission powers $\bm{\rho}$ for a given $\bm{V}$ and $\bm{\Theta}$. Similar to the method used for updating $\bm{\Theta}$, we refine the estimate of BS transmission powers by following the gradient direction while ensuring that all BSs satisfy the upper bound $\rho_n \leq \rho_{\max}$. To achieve this, we use the projected gradient ascent on the feasible region $\bm{\Lambda}$.  The partial derivatives are provided next.

\begin{Proposition}\label{power_gradient_sumlograte_outage_function}
    The partial derivative of the performance function $\mathcal{P}^{(1)}_{\gamma_1}(\bm{V}, \bm{\Theta}, \bm{\rho})$ in (\ref{SumLogRate_App1_performance_function}) w.r.t. the BS transmission power $\rho_n$ is given by:
\begin{align}\label{gamma_1_1_partial_derivative_wrt_power}
    &\frac{\partial}{\partial \rho_n} \mathcal{P}^{(1)}_{\gamma_1} = \sum_{m=1}^{N} \int_{V_m} \Bigg [ \frac{\beta \cdot \log_2(e) \cdot \log_2(e) \cdot \ln{10} \cdot 0.1}{\log_2\big(1 + {\sf SINR_{lin}^{(m)}}(\bm{q})\big)} \nonumber\\& \times \frac{{\sf SINR_{lin}^{(m)}}(\bm{q})}{1 + {\sf SINR_{lin}^{(m)}}(\bm{q})}  + \kappa (1 - \beta) \cdot \sigma\Big(\kappa \big[{\sf SINR_{dB}^{(m)}}(\bm{q}) - T \big] \Big) \nonumber\\& \times\! \bigg[1 - \sigma\Big(\kappa \big[{\sf SINR_{dB}^{(m)}}(\bm{q}) - T \big] \Big)\bigg] \Bigg ] \!\cdot\! \frac{\partial}{\partial \rho_n} {\sf SINR_{dB}^{(m)}}(\bm{q}) \lambda(\bm{q})d\bm{q},
\end{align}
where
\begin{equation}\label{partial_sinr_rho_derivatives}
    \frac{\partial {\sf SINR_{dB}^{(m)}}(\bm{q})}{\partial \rho_n} = 
\begin{cases}
    1,                                     &\text{if } m = n \\
    - \frac{ {\sf RSS_{lin}^{(n)}}(\bm{q}) \cdot {\sf SINR_{lin}^{(m)}}(\bm{q}) }{{\sf RSS_{lin}^{(m)}}(\bm{q}) },            &\text{o.w.} 
\end{cases}  
\end{equation}
and $\sigma(x) = \frac{1}{1 + e^{-x}}$ is the sigmoid function.
\end{Proposition}

\noindent The proof of Proposition \ref{power_gradient_sumlograte_outage_function} is provided in Appendix \ref{Appendix_C}.

\subsubsection*{Optimization algorithm}

Using Propositions \ref{optimal_cell_partitioning_sum_log_rate}, \ref{theta_gradient_sumlograte_outage_function}, and \ref{power_gradient_sumlograte_outage_function}, we design the following optimization algorithm.

\begin{Algorithm}\label{gamma_1_1_algorithm}
The optimization of transmit power and vertical tilts for a coverage-capacity trade-off starts with a random initialization of tilts and powers and iterates between the following three steps:
\begin{itemize}
    \item For the given $\bm{\Theta}$ and $\bm{\rho}$, update the cell partitioning $\bm{V}$ according to (\ref{optimal_cell_partitioning_sum_log_rate_eq});
    \item Using the partial derivatives in (\ref{gamma_1_1_partial_derivative_wrt_theta}), update the BS vertical antenna tilts $\bm{\Theta}$ via gradient ascent while $\bm{V}$ and $\bm{\rho}$ are fixed;
    \item Using the partial derivatives in (\ref{gamma_1_1_partial_derivative_wrt_power}), update the BS transmission powers $\bm{\rho}$ via projected gradient ascent with the feasible region $\bm{\Lambda}$ while $\bm{V}$ and $\bm{\Theta}$ are fixed. 
\end{itemize}
\end{Algorithm}

\begin{Proposition}\label{convergence_gamma_1_1_algorithm}
    Algorithm~\ref{gamma_1_1_algorithm} is an iterative improvement algorithm and converges.
\end{Proposition}

\noindent The proof of Proposition \ref{convergence_gamma_1_1_algorithm} is provided in Appendix \ref{Appendix_D}.


\subsection{Application \#2: Base Station Deployment Optimization for Coverage-capacity Trade-off}\label{sum_log_rate_application_2_section}

For the coverage-capacity trade-off, KPI $\gamma_1^{(n)}$ in (\ref{second_KPI}), the maximization of (\ref{general_objective_with_deployment}) becomes:
\begin{align}\label{optimization_problem_KPI1_App2}
&\max_{\substack{\bm{V}, \bm{\Theta}, \bm{\rho}, \overline{\widehat{\bm{P}}}, \overline{\widehat{\bm{\Phi}}} \\ \bm{\rho}\in \bm{\Lambda} } } \mathcal{P}^{(2)}_{\gamma_1}\big(\bm{V}, \bm{\Theta}, \bm{\rho}, \overline{\widehat{\bm{P}}}, \overline{\widehat{\bm{\Phi}}}\big)  \nonumber\\ 
=&\max_{\substack{\bm{V}, \bm{\Theta}, \bm{\rho}, \overline{\widehat{\bm{P}}}, \overline{\widehat{\bm{\Phi}}} \\ \bm{\rho}\in \bm{\Lambda} } }  \sum_{n=1}^{N} \int_{V_n} \gamma_{1}^{(n)}\big(\bm{q}; \bm{\Theta}, \bm{\rho}, \overline{\widehat{\bm{P}}}, \overline{\widehat{\bm{\Phi}}}\big) \lambda(\bm{q})d\bm{q},
\end{align}
where $\bm{\Lambda} = (-\infty, \rho_{\max}]^{N}$ is the feasible region for the vector of transmission powers $\bm{\rho}$. 

The goal of (\ref{optimization_problem_KPI1_App2}) is to optimize the location and the reference bearing for each new site, in addition to optimizing the cell partitioning, vertical antenna tilts, and transmission powers for all base stations. Similar to the approach in Section \ref{sum_log_rate_application_1_section}, we aim to design an algorithm that iteratively optimizes each of the five parameters $\bm{V}$, $\bm{\Theta}$, $\bm{\rho}$,  $\overline{\widehat{\bm{P}}}$, and $\overline{\widehat{\bm{\Phi}}}$ while keeping the other four fixed. 

\subsubsection*{Optimal cell partitioning}

Similar to the proof of Proposition \ref{optimal_cell_partitioning_sum_log_rate}, it can be shown that the optimal cell partitioning $\bm{V}$ that maximizes the performance function $\mathcal{P}^{(2)}_{\gamma_1}$ in (\ref{optimization_problem_KPI1_App2}) is given by (\ref{optimal_cell_partitioning_sum_log_rate_eq}), i.e., each user location $\bm{q}$ is associated with the base station that provides the highest RSS.

\subsubsection*{Optimal vertical antenna tilts and transmission powers}

Optimization for antenna tilts and transmission powers proceeds using the gradient ascent and the projected gradient ascent, respectively. The partial derivatives of $\mathcal{P}^{(2)}_{\gamma_1}$ w.r.t. $\theta_n$ and $\rho_n$ have similar expressions as the partial derivatives of $\mathcal{P}^{(1)}_{\gamma_1}$ in Propositions \ref{theta_gradient_sumlograte_outage_function} and \ref{power_gradient_sumlograte_outage_function},  respectively. 

We now employ gradient ascent to optimize the deployment and the reference bearing of new BS site locations. 

\subsubsection*{Optimal locations of new deployment sites}

The gradient of the performance function $\mathcal{P}^{(2)}_{\gamma_1}$ w.r.t. site location $\widehat{\bm{p}}_m$ is as follows. 

\begin{Proposition}\label{deployment_gradient_sumlograte_outage_function}
    The gradient of the performance function $\mathcal{P}^{(2)}_{\gamma_1}\big(\bm{V}, \bm{\Theta}, \bm{\rho}, \overline{\widehat{\bm{P}}}, \overline{\widehat{\bm{\Phi}}}\big)$ in (\ref{optimization_problem_KPI1_App2}) w.r.t. the $m$-th site location is:
\begin{align}\label{deployment_gradient_sumlograte_outage_function_eq}
    &\nabla_{\widehat{\bm{p}}_m} \mathcal{P}^{(2)}_{\gamma_1} = \sum_{n=1}^{N} \int_{V_n} \Bigg [ \frac{\beta \cdot \log_2(e) \cdot \log_2(e) \cdot \ln{10} \cdot 0.1}{\log_2\big(1 + {\sf SINR_{lin}^{(n)}}(\bm{q})\big)} \nonumber\\
    & \times \frac{{\sf SINR_{lin}^{(n)}}(\bm{q})}{1 + {\sf SINR_{lin}^{(n)}}(\bm{q})}  + \kappa (1 - \beta) \cdot \sigma\Big(\kappa \big[{\sf SINR_{dB}^{(n)}}(\bm{q}) - T \big] \Big) \nonumber\\
    & \times\! \bigg[1 - \sigma\Big(\kappa \big[{\sf SINR_{dB}^{(n)}}(\bm{q}) - T \big] \Big)\bigg] \Bigg ] \!\cdot \nabla_{\bm{\widehat{p}}_m} {\sf SINR_{dB}^{(n)}}(\bm{q}) \lambda(\bm{q})d\bm{q}.    
\end{align}    
If $n \notin \{3m-2, 3m-1, 3m\}$, the term $\nabla_{\bm{\widehat{p}}_m} {\sf SINR_{dB}^{(n)}}(\bm{q})$ is:
\begin{align}\label{eq-59_prop}
    \nabla_{\widehat{\bm{p}}_m} {\sf SINR_{dB}^{(n)}} & =    - \frac{ {\sf SINR_{lin}^{(n)}} }{ {\sf RSS_{lin}^{(n)}}   } \!\times\! \Big [ {\sf RSS_{lin}^{(3m-2)}} \!\cdot\! \nabla_{\bm{p}_{3m-2}} {\sf RSS_{dBm}^{(3m-2)}} \nonumber\\& + {\sf RSS_{lin}^{(3m-1)}} \cdot \nabla_{\bm{p}_{3m-1}} {\sf RSS_{dBm}^{(3m-1)}} \nonumber\\& + {\sf RSS_{lin}^{(3m)}} \cdot \nabla_{\bm{p}_{3m}} {\sf RSS_{dBm}^{(3m)}}  \Big ].    
\end{align}
However, if $n \in \{3m-2, 3m-1, 3m\}$, we have:
\begin{align}\label{eq-60_prop}
    & \nabla_{\widehat{\bm{p}}_m} {\sf SINR_{dB}^{(n)}} = \nabla_{\bm{p}_n} {\sf RSS_{dBm}^{(n)}} - \nonumber\\& \frac{ \Big[ {\sf RSS_{lin}^{(n')}} \cdot \nabla_{\bm{p}_{n'}} {\sf RSS_{dBm}^{(n')}}  +  {\sf RSS_{lin}^{(n'')}} \cdot \nabla_{\bm{p}_{n''}} {\sf RSS_{dBm}^{(n'')}} \Big]}{ \Big [ \sum_{j \neq n} {\sf RSS_{lin}^{(j)}} + \sigma_{\sf lin}^2 \Big ]  },
\end{align}
where $\{n, n', n''\} = \{3m-2, 3m-1, 3m \}$. For each $n$, the term $\nabla_{\bm{p}_{n}} {\sf RSS_{dBm}^{(n)}}$ is given by:
\begin{align}\label{RSS_grad_pk}
    &\nabla_{\bm{p}_n} {\sf RSS_{dBm}^{(n)}}(\bm{q}; \theta_n, \rho_n, \bm{p}_n, \phi_n) = \nonumber\\& \frac{-24}{\theta_{{\sf 3dB}}^{2}} (\theta_{n,\bm{q}} - \theta_n) \cdot \frac{(h_{n,B} - h_{\bm{q}}) \cdot (\bm{p}_n - \bm{q})}{\|\bm{p}_n - \bm{q} \|^3 + \|\bm{p}_n - \bm{q}\| (h_{n,B} - h_{\bm{q}})^2      } \nonumber \\ & - \frac{24}{\phi^2_{{\sf 3dB}}} (\phi_{n,\bm{q}} - \phi_n) \cdot \bigg(\frac{q_y - p_{n,y}}{\|\bm{q} - \bm{p}_n\|^2}, \frac{p_{n,x} - q_x}{\|\bm{q} - \bm{p}_n\|^2} \bigg) \nonumber\\& - \frac{b_{\bm{q}} \log_{10}(e)\cdot (\bm{p}_n - \bm{q})}{\|\bm{q} - \bm{p}_n \|^2 + (h_{\bm{q}} - h_{n,B})^2 }.
\end{align}
\end{Proposition}

\noindent 
{The proof of Proposition \ref{deployment_gradient_sumlograte_outage_function} is similar to those of Propositions \ref{theta_gradient_sumlograte_outage_function} and \ref{power_gradient_sumlograte_outage_function} and is omitted because of the page limit.}

\subsubsection*{Optimal bearings of new deployment sites}

The gradient of the performance function $\mathcal{P}^{(2)}_{\gamma_1}$ w.r.t. the reference bearing of the new BS sites is as follows. 

\begin{Proposition}\label{azimuth_gradient_sumlograte_outage_function}
    The partial derivatives of the performance function  $\mathcal{P}^{(2)}_{\gamma_1}\big(\bm{V}, \bm{\Theta}, \bm{\rho}, \overline{\widehat{\bm{P}}}, \overline{\widehat{\bm{\Phi}}}\big)$ in (\ref{optimization_problem_KPI1_App2}) w.r.t. the site $m$'s reference bearing is:
\begin{align}\label{azimuth_gradient_sumlograte_outage_function_eq}
    &\frac{\partial \mathcal{P}^{(2)}_{\gamma_1}}{\partial \widehat{\phi}_m}  = \sum_{n=1}^{N} \int_{V_n} \Bigg [ \frac{\beta \cdot \log_2(e) \cdot \log_2(e) \cdot \ln{10} \cdot 0.1}{\log_2\big(1 + {\sf SINR_{lin}^{(n)}}(\bm{q})\big)} \nonumber\\
    &\times \frac{{\sf SINR_{lin}^{(n)}}(\bm{q})}{1 + {\sf SINR_{lin}^{(n)}}(\bm{q})}  + \kappa (1 - \beta) \cdot \sigma\Big(\kappa \big[{\sf SINR_{dB}^{(n)}}(\bm{q}) - T \big] \Big) \nonumber\\
    & \times \bigg[1 - \sigma\Big(\kappa \big[{\sf SINR_{dB}^{(n)}}(\bm{q}) - T \big] \Big)\bigg] \Bigg ] \cdot \frac{\partial  {\sf SINR_{dB}^{(n)}}(\bm{q})}{\partial \widehat{\phi}_m} \lambda(\bm{q})d\bm{q}.    
\end{align}      
If $n \notin \{3m-2, 3m-1, 3m\}$, the term $\frac{\partial  {\sf SINR_{dB}^{(n)}}(\bm{q})}{\partial \widehat{\phi}_m}$ is:
\begin{align}\label{eq-61_proposition}
    &\frac{\partial {\sf SINR_{dB}^{(n)}} }{\partial \widehat{\phi}_m } = - \frac{ {\sf SINR_{lin}^{(n)}} }{  {\sf RSS_{lin}^{(n)}} } \cdot \frac{24}{\phi^2_{\textrm{3dB}} } \cdot\!\!  \sum_{t=3m-2}^{3m}  {\sf RSS_{lin}^{(t)}} \cdot \left ( \phi_{t, \bm{q}} - \phi_{t} \right).
\end{align}
However, if $n \in \{3m-2, 3m-1, 3m\}$, we have:
\begin{align}\label{eq-62_proposition}
    & \frac{\partial {\sf SINR_{dB}^{(n)}} }{\partial  \widehat{\phi}_m } = \frac{ 24 }{ \phi_{\textrm{3dB}}^2 } \bigg[ \left ( \phi_{n, \bm{q}} - \phi_n \right) -   \frac{ {\sf SINR_{lin}^{(n)}} }{  {\sf RSS_{lin}^{(n)}} } \times \nonumber\\&  \Big( {\sf RSS_{lin}^{(n')}} \cdot \left( \phi_{n', \bm{q}} - \phi_{n'} \right)  +  {\sf RSS_{lin}^{(n'')}} \cdot \left( \phi_{n'', \bm{q}} - \phi_{n''} \right)  \Big) \bigg ], 
\end{align}
where $\{n, n', n''\} = \{3m-2, 3m-1, 3m \}$.
\end{Proposition}

\noindent 
{The proof of Proposition \ref{azimuth_gradient_sumlograte_outage_function} is similar to that of Proposition \ref{theta_gradient_sumlograte_outage_function} and is omitted because of the page limit.}

\subsubsection*{Optimization algorithm}

Using Propositions \ref{deployment_gradient_sumlograte_outage_function} and \ref{azimuth_gradient_sumlograte_outage_function}, we design the following optimization algorithm.


\begin{Algorithm}\label{gamma_1_2_algorithm}
The BS deployment optimization for a coverage-capacity trade-off starts with a random initialization of the parameters and proceeds as follows:
\begin{itemize}
    \item Given $\bm{\Theta}$, $\bm{\rho}$, $\overline{\widehat{\bm{P}}}$, and $\overline{\widehat{\bm{\Phi}}}$, optimize the cell partitioning $\bm{V}$ by assigning each user $\bm{q}$ to the base station that provides the highest RSS value at that user location;
    \item Apply gradient ascent to optimize the vertical antenna tilts $\bm{\Theta}$ for a given set of $\bm{V}$, $\bm{\rho}$, $\overline{\widehat{\bm{P}}}$, and $\overline{\widehat{\bm{\Phi}}}$ values;
    \item Apply the projected gradient ascent with the feasible region $\bm{\Lambda}$ to optimize the BS transmission powers $\bm{\rho}$ for a given set of $\bm{V}$, $\bm{\Theta}$, $\overline{\widehat{\bm{P}}}$, and $\overline{\widehat{\bm{\Phi}}}$ values;
    \item Given $\bm{V}$, $\bm{\Theta}$, $\bm{\rho}$, and $\overline{\widehat{\bm{\Phi}}}$, apply gradient ascent to optimize site locations using gradient expressions in (\ref{deployment_gradient_sumlograte_outage_function_eq});
    \item Given  $\bm{V}$, $\bm{\Theta}$, $\bm{\rho}$, and $\overline{\widehat{\bm{P}}}$, apply gradient ascent to optimize sites' reference azimuth orientations using the partial derivatives calculated in (\ref{azimuth_gradient_sumlograte_outage_function_eq}).
\end{itemize}
The algorithm iterates between the above five steps until the convergence criterion is met.
\end{Algorithm}

\begin{Proposition}\label{convergence_gamma_1_2_algorithm}
    Algorithm~\ref{gamma_1_2_algorithm} is an iterative improvement algorithm and converges.
\end{Proposition}
\noindent The proof of Proposition \ref{convergence_gamma_1_2_algorithm} is similar to that of Proposition \ref{convergence_gamma_1_1_algorithm} and is omitted.


\section{Optimizing Cellular Networks for\\Maximum Capacity per Region}\label{capacity_per_region_section}

In this section, we study the two applications introduced in Sections \ref{Tilt_Optimization} and \ref{Deployment_Optimization} for the capacity-per-region KPI $\gamma_2^{(n)}$ given in (\ref{first_KPI_biased_capacity_per_region}).


\subsection{Application \#1: Base Station Antenna Tilt Optimization and Power Allocation for Maximum Capacity per Region}\label{cap_per_region_application_1}

For the capacity-per-region KPI $\gamma_2^{(n)}$ in (\ref{first_KPI_biased_capacity_per_region}), the maximization of (\ref{general_objective_no_deployment}) becomes:
\begin{align}\label{general_objective_no_deployment_first_KPI}
& \max_{\substack{\bm{V}, \bm{\Theta}, \bm{\rho} \\ \bm{\rho}\in \bm{\Lambda} } } \mathcal{P}^{(1)}_{\gamma_2}(\bm{V}, \bm{\Theta}, \bm{\rho}) 
\nonumber\\
=& \max_{\substack{\bm{V}, \bm{\Theta}, \bm{\rho} \\ \bm{\rho}\in \bm{\Lambda} } } \sum_{n=1}^{N} \int_{V_n} \frac{\log_2 \big(1 + {\sf SINR_{lin}^{(n)}}(\bm{q}; \bm{\Theta}, \bm{\rho}) \big)}{o_n + \int_{V_n} \lambda(\bm{q})d\bm{q}} \lambda(\bm{q})d\bm{q},
\end{align}
where $\bm{\Lambda} = (-\infty, \rho_{\max}]^N$ is the feasible region for the BS transmission powers $\bm{\rho}$. 

The goal of (\ref{general_objective_no_deployment_first_KPI}) is to maximize the capacity per region with respect to the cell partitioning $\bm{V}$, the BS vertical antenna tilts $\bm{\Theta}$, and the BS transmission powers $\bm{\rho}$. Our approach to solve (\ref{general_objective_no_deployment_first_KPI}) 
is similar to the one adopted to tackle (\ref{SumLogRate_App1_performance_function}) in Section~\ref{sum_log_rate_application_1_section} and detailed as follows.


\subsubsection*{Optimal cell partitioning}

First, we state the following.  
\begin{Lemma}\label{RSS_optimal_for_rate}
    We define the auxiliary performance function 
\begin{align}\label{auxiliary_objective_no_deployment_first_KPI}
\mathcal{P}^{(1)}_{\overline{\gamma}_2}(\bm{V}, \bm{\Theta}, \bm{\rho}) &= \sum_{n=1}^{N} \int_{V_n} \overline{\gamma}_2^{(n)}(\bm{q}; \bm{\Theta}, \bm{\rho}) \lambda(\bm{q})d\bm{q} ,
\end{align}
where the auxiliary KPI $\overline{\gamma}_2^{(n)}(\bm{q}; \bm{\Theta}, \bm{\rho})$ is defined as:
\begin{equation}\label{auxiliary_first_KPI_biased_capacity_per_region}
    \overline{\gamma}_2^{(n)}(\bm{q}; \bm{\Theta}, \bm{\rho}) = \log_2 \big(1 + {\sf SINR_{lin}^{(n)}}(\bm{q}; \bm{\Theta}, \bm{\rho}) \big).
\end{equation}
Then, the optimal cell partitioning $\bm{V}^*$ that maximizes $\mathcal{P}^{(1)}_{\overline{\gamma}_2}(\bm{V}, \bm{\Theta}, \bm{\rho})$ for a given $\bm{\Theta}$ and $\bm{\rho}$ is given by:
\begin{align}\label{optimal_cell_partitioning_auxiliary_KPI}
    V^*_n = \{\bm{q} \in Q \mid {\sf RSS_{dBm}^{(n)}}(\bm{q}) \geq {\sf RSS_{dBm}^{(k)}}(\bm{q}), \textrm{ } \forall k \neq n  \},
\end{align}
for each $n \in \{1, \cdots, N\}$.

\textit{Proof. } The proof follows directly from Lemma \ref{ascending_func_of_sinr}  since $f(x) = \log_2(1 + x)$ is a continuous and increasing function. \qed
\end{Lemma}

\noindent 
Lemma \ref{RSS_optimal_for_rate} indicates that in the absence of the denominator in (\ref{first_KPI_biased_capacity_per_region}), the optimal cell partitioning is given by (\ref{optimal_cell_partitioning_auxiliary_KPI}). 
Inspired by this observation, we update the cell partitioning for any given network configuration according to (\ref{optimal_cell_partitioning_auxiliary_KPI}) if and only if it improves the KPI for that particular configuration.

\subsubsection*{Optimal vertical antenna tilts}

Optimization for BS vertical antenna tilts $\bm{\Theta}$ is carried out via the gradient ascent method in which partial derivatives are expressed as follows.

\begin{Proposition}\label{application_1_KPI_1_derivative_wrt_theta}
The partial derivative of the performance function $\mathcal{P}^{(1)}_{\gamma_2}(\bm{V}, \bm{\Theta}, \bm{\rho})$ w.r.t. $\theta_n$ is given by:
\begin{align}\label{application_1_KPI_1_derivative_wrt_theta_eq}
    &\frac{\partial}{\partial \theta_n} \mathcal{P}^{(1)}_{\gamma_2} = \frac{\log_2(e) \cdot \ln(10) \cdot 0.1 }{o_n + \int_{V_n} \lambda(\bm{q})d\bm{q} } \nonumber\\& \times \int_{V_n} \frac{ {\sf SINR_{lin}^{(n)}}(\bm{q}) }{ 1 + {\sf SINR_{lin}^{(n)}}(\bm{q}) } \cdot \frac{24}{\theta^2_{\sf 3dB}} (\theta_{n,\bm{q}} - \theta_n) \lambda(\bm{q})d\bm{q}  \nonumber\\& -\!\!
    \sum_{m=1, m\neq n}^{N} \frac{\log_2(e) \cdot \ln(10) \cdot 0.1 }{o_m + \int_{V_m} \lambda(\bm{q})d\bm{q} }  \times \int_{V_m} \frac{ {\sf SINR_{lin}^{(m)}}(\bm{q}) }{ 1 + {\sf SINR_{lin}^{(m)}}(\bm{q}) } \times \nonumber\\& \frac{\frac{24}{\theta^2_{\sf 3dB}}\cdot (\theta_{n,\bm{q}} - \theta_n)\cdot {\sf RSS_{lin}^{(n)}}(\bm{q}) \cdot {\sf SINR_{lin}^{(m)}} (\bm{q}) }{{\sf RSS_{lin}^{(m)}} (\bm{q}) } \lambda(\bm{q})d\bm{q}.
\end{align}
\end{Proposition}

\noindent 
{The proof of Proposition \ref{application_1_KPI_1_derivative_wrt_theta} is similar to that of Proposition \ref{theta_gradient_sumlograte_outage_function} and is omitted because of the page limit.}

\subsubsection*{Optimal transmission powers}

Finally, the optimization for BS transmission powers $\bm{\rho}$ is carried out using the projected gradient ascent method with the feasible region $\bm{\Lambda}$. The partial derivatives needed for this purpose are provided next.

\begin{Proposition}\label{application_1_KPI_1_derivative_wrt_power}
The partial derivative of the performance function $\mathcal{P}^{(1)}_{\gamma_2}(\bm{V}, \bm{\Theta}, \bm{\rho})$ w.r.t. $\rho_n$ is given by:

\begin{align}\label{application_1_KPI_1_derivative_wrt_power_eq}
    &\frac{\partial}{\partial \rho_n} \mathcal{P}^{(1)}_{\gamma_2} = \frac{\log_2(e) \cdot \ln(10) \cdot 0.1 }{o_n + \int_{V_n} \lambda(\bm{q})d\bm{q} }  \nonumber\\& \times \int_{V_n} \frac{ {\sf SINR_{lin}^{(n)}}(\bm{q}) }{ 1 + {\sf SINR_{lin}^{(n)}}(\bm{q}) } \cdot 1 \cdot \lambda(\bm{q})d\bm{q} \nonumber\\&-
     \sum_{m=1, m\neq n}^{N} \frac{\log_2(e) \cdot \ln(10) \cdot 0.1 }{o_m + \int_{V_m} \lambda(\bm{q})d\bm{q} } \nonumber\\& \times \int_{V_m} \frac{ {\sf SINR_{lin}^{(m)}}(\bm{q}) }{ 1 + {\sf SINR_{lin}^{(m)}}(\bm{q}) } \cdot \frac{ {\sf RSS_{lin}^{(n)}}(\bm{q}) \cdot {\sf SINR_{lin}^{(m)}}(\bm{q}) }{{\sf RSS_{lin}^{(m)}}(\bm{q}) } \cdot \lambda(\bm{q})d\bm{q}.
\end{align}
\end{Proposition}

\noindent 
{The proof of Proposition \ref{application_1_KPI_1_derivative_wrt_power} is similar to that of Proposition  \ref{power_gradient_sumlograte_outage_function} and is omitted because of the page limit.}

\subsubsection*{Optimization algorithm}

Using Propositions \ref{application_1_KPI_1_derivative_wrt_theta} and \ref{application_1_KPI_1_derivative_wrt_power}, and Lemma~\ref{RSS_optimal_for_rate} for updating the cell partitioning, we design the following optimization algorithm.

\begin{Algorithm}\label{gamma_2_1_algorithm}
The optimization of transmit power and vertical tilts for maximum capacity per area starts with a random initialization of tilts $\bm{\Theta}$ and powers $\bm{\rho}$, and iterates between the following three steps until the convergence criterion is met:
\begin{itemize}
    \item For the given $\bm{\Theta}$ and $\bm{\rho}$, update the cell partitioning $\bm{V}$ according to Lemma~\ref{RSS_optimal_for_rate} if and only if KPI is improved;
    \item Using the partial derivatives in \eqref{application_1_KPI_1_derivative_wrt_theta_eq}, update the BS vertical antenna tilts $\bm{\Theta}$ via gradient ascend while $\bm{V}$ and $\bm{\rho}$ are fixed;
    \item Using the partial derivatives in \eqref{application_1_KPI_1_derivative_wrt_power_eq}, update the BS transmission powers $\bm{\rho}$ via projected gradient ascent with the feasible region $\bm{\Lambda} = (-\infty, \rho_{\max}]^N$ while $\bm{V}$ and $\bm{\Theta}$ are held fixed. 
\end{itemize}
\end{Algorithm}

\begin{Proposition}\label{cap_per_reg_application1_alg_convergence}
    Algorithm~\ref{gamma_2_1_algorithm} is an iterative improvement algorithm and converges.
\end{Proposition}

\noindent 
{The proof of Proposition \ref{cap_per_reg_application1_alg_convergence} is similar to that of Proposition \ref{convergence_gamma_1_1_algorithm} and is omitted because of the page limit.}


\subsection{Application \#2: Base Station Deployment Optimization for Maximum Capacity per Region}\label{cap_per_region_application_2}

For the capacity-per-region KPI $\gamma_2^{(n)}$ in (\ref{first_KPI_biased_capacity_per_region}), the maximization of (\ref{general_objective_with_deployment}) becomes:
\begin{align}\label{cap_per_region_application_2_general_objective}
& \max_{\substack{\bm{V}, \bm{\Theta}, \bm{\rho}, \overline{\widehat{\bm{P}}}, \overline{\widehat{\bm{\Phi}}} \\ \bm{\rho}\in \bm{\Lambda} } } \mathcal{P}^{(2)}_{\gamma_2}(\bm{V}, \bm{\Theta}, \bm{\rho}, \overline{\widehat{\bm{P}}}, \overline{\widehat{\bm{\Phi}}}) \nonumber\\
& = \!\!\!\!\!\!\!\max_{\substack{\bm{V}, \bm{\Theta}, \bm{\rho}, \overline{\widehat{\bm{P}}}, \overline{\widehat{\bm{\Phi}}} \\ \bm{\rho}\in \bm{\Lambda} } }   \sum_{n=1}^{N} \! \int_{V_n} \!\!\! \frac{\log_2\! \big(1 \!+\! {\sf SINR_{lin}^{(n)}}(\bm{q}; \bm{\Theta}, \bm{\rho}, \overline{\widehat{\bm{P}}}, \overline{\widehat{\bm{\Phi}}}) \big)}{o_n + \int_{V_n} \lambda(\bm{q})d\bm{q}} \lambda(\bm{q})d\bm{q},
\end{align}
where $\bm{\Lambda} = (-\infty, \rho_{\max}]^{N}$ is the feasible set for $\bm{\rho}$. 

The goal of (\ref{cap_per_region_application_2_general_objective}) is to optimize the location and the reference bearing for each new site, in addition to optimizing the cell partitioning, vertical antenna tilts, and transmission powers for all base stations. Our approach to solve (\ref{cap_per_region_application_2_general_objective}) 
is similar to the one adopted to tackle (\ref{general_objective_no_deployment_first_KPI}) in Section~\ref{cap_per_region_application_1}, with an extra optimization carried out over the position and orientation of new BSs.

To solve this new optimization problem, similar to the way that we extended the results of Section \ref{sum_log_rate_application_1_section} to those of Section \ref{sum_log_rate_application_2_section}, we only need to calculate the new gradients w.r.t. the site location $\widehat{\bm{p}}_m$ and the reference bearing $\widehat{\phi}_m$. The approaches to calculate these gradients, the resulting iterative improvement
algorithm, and the proof of the algorithm convergence are very similar to previous propositions. As such, in what follows, we only present them without proving them to archive the formulas. 

\begin{Proposition}\label{application_2_KPI_1_derivative_wrt_locations}
The gradient of the performance function $\mathcal{P}^{(2)}_{\gamma_2}(\bm{V}, \bm{\Theta}, \bm{\rho}, \overline{\widehat{\bm{P}}}, \overline{\widehat{\bm{\Phi}}})$ w.r.t. the site location $\widehat{\bm{p}}_m$ is given by:
\begin{align}\label{application_2_KPI_1_derivative_wrt_locations_eq}
    &\nabla_{\widehat{\bm{p}}_m} \mathcal{P}^{(2)}_{\gamma_2} = \sum_{n=1}^{N} \frac{\log_2(e) \cdot \ln(10) \cdot 0.1 }{o_n + \int_{V_n} \lambda(\bm{q})d\bm{q} } \nonumber\\& \times \int_{V_n} \frac{ {\sf SINR_{lin}^{(n)}}(\bm{q}) }{ 1 + {\sf SINR_{lin}^{(n)}}(\bm{q}) } \cdot \nabla_{\widehat{\bm{p}}_m} {\sf SINR_{dB}^{(n)}}(\bm{q}) \lambda(\bm{q})d\bm{q},    
\end{align}   
where depending on the values of $m$ and $n$, the term $\nabla_{\widehat{\bm{p}}_m} {\sf SINR_{dB}^{(n)}}$ is given by (\ref{eq-59_prop}) and (\ref{eq-60_prop}).
\end{Proposition}





\begin{Proposition}\label{application_2_KPI_1_derivative_wrt_azimuth}
The partial derivative of the performance function $\mathcal{P}^{(2)}_{\gamma_2}(\bm{V}, \bm{\Theta}, \bm{\rho}, \overline{\widehat{\bm{P}}}, \overline{\widehat{\bm{\Phi}}})$ w.r.t. $\widehat{\phi}_m$ is given by:
\begin{align}\label{total_grad_wrt_phi}
    &\frac{\partial}{\partial \widehat{\phi}_m} \mathcal{P}^{(2)}_{\gamma_2} = \sum_{n=1}^{N} \frac{\log_2(e) \cdot \ln(10) \cdot 0.1 }{o_n + \int_{V_n} \lambda(\bm{q})d\bm{q} } \nonumber\\& \times \int_{V_n} \frac{ {\sf SINR_{lin}^{(n)}}(\bm{q}) }{ 1 + {\sf SINR_{lin}^{(n)}}(\bm{q}) } \cdot \frac{\partial}{\partial \widehat{\phi}_m} {\sf SINR_{dB}^{(n)}}(\bm{q}) \lambda(\bm{q})d\bm{q},
\end{align}
where depending on the values of $m$ and $n$, the term $\frac{\partial {\sf SINR_{dB}^{(n)}} }{\partial \widehat{\phi}_m}$ is given by (\ref{eq-61_proposition}) or (\ref{eq-62_proposition}).
\end{Proposition}






\begin{Algorithm}\label{gamma_2_2_algorithm}
The BS deployment optimization for maximum capacity per area starts with a random initialization of tilts $\bm{\Theta}$,  powers $\bm{\rho}$, site locations $\overline{\widehat{\bm{P}}}$, and reference bearings $\overline{\widehat{\bm{\Phi}}}$, and iterates between the following steps until the convergence criterion is met:
\begin{itemize}
    \item For the given $\bm{\Theta}$, $\bm{\rho}$, $\overline{\widehat{\bm{P}}}$, and $\overline{\widehat{\bm{\Phi}}}$, update the cell partitioning $\bm{V}$ according to Lemma~\ref{RSS_optimal_for_rate} if and only if it improves the KPI;
    \item Using the partial derivatives in \eqref{application_1_KPI_1_derivative_wrt_theta_eq}, update the BS vertical antenna tilts $\bm{\Theta}$ via gradient ascend while $\bm{V}$, $\bm{\rho}$, $\overline{\widehat{\bm{P}}}$, and $\overline{\widehat{\bm{\Phi}}}$ are fixed;
    \item Using the partial derivatives in \eqref{application_1_KPI_1_derivative_wrt_power_eq}, update the BS transmission powers $\bm{\rho}$ via projected gradient ascend with the feasible region $\bm{\Lambda} = (-\infty, \rho_{\max}]^{N}$ for a given $\bm{V}$, $\bm{\Theta}$, $\overline{\widehat{\bm{P}}}$, and $\overline{\widehat{\bm{\Phi}}}$; 
    \item Update the site locations $\overline{\widehat{\bm{P}}}$ via gradient ascent while $\bm{V}$, $\bm{\Theta}$, $\bm{\rho}$, and $\overline{\widehat{\bm{\Phi}}}$ are fixed using gradient expressions in \eqref{application_2_KPI_1_derivative_wrt_locations_eq};
    \item For a given $\bm{V}$, $\bm{\Theta}$, $\bm{\rho}$, and $\overline{\widehat{\bm{P}}}$, update the sites' reference bearings $\overline{\widehat{\bm{\Phi}}}$ via gradient ascent using the partial derivatives calculated in \eqref{total_grad_wrt_phi}.    
\end{itemize}
\end{Algorithm}



\section{Case Study}\label{case-study}

Robust connectivity is crucial for applications like drone delivery services and advanced urban air mobility. Traditional terrestrial cellular BSs are optimized for 2D ground-level connectivity, and achieving 3D connectivity may require re-engineering the cellular network \cite{GerLopBen2022,MozLinHay2021,KanMezLoz2021}. With UAVs expected to operate within designated aerial paths or corridors, we focus on optimizing connectivity within these areas. Recent studies, with the exception of our work in \cite{KarGerJaf2023}, have approached this problem through ad-hoc system-level optimizations of simplified setups \cite{bernabe2022optimization, bernabe2023novel, maeng2023base, chowdhury2021ensuring, singh2021placement}. However, a general mathematical framework for analyzing and designing cellular networks for both ground users and UAV corridors is still needed.  In this section, we demonstrate the effectiveness of our mathematical framework by jointly optimizing (i) the vertical antenna tilts and the transmission powers of all BSs and (ii) the locations and bearings of newly deployed BSs to enhance the connectivity performance of ground and aerial users. We begin by describing our network setup.

\begin{table}[t!]
\centering
\caption{ Coverage-capacity performance comparison with and without the deployment of new BSs. Values are reported for $r = 0.5$ and two choices of distributions for GUEs. }
\begin{tabular}{|cc||ccc|}
 \toprule
\backslashbox{Algorithm}{GUE Distribution}          &\!\!\!\!\!\!&  Uniform  &\!\!\!\!\!\!&  Gaussian Mixture  \\\hline \hline
{\bf {\small Algorithm \ref{gamma_1_1_algorithm}}}  &\!\!\!\!\!\!&  $1.2598$ &\!\!\!\!\!\!&   $1.3072$         \\ 
{\bf {\small Algorithm \ref{gamma_1_2_algorithm}}}  &\!\!\!\!\!\!&  $1.3443$ &\!\!\!\!\!\!&   $1.3785$         \\ 
 \bottomrule 
\end{tabular}
  \label{coverage_capacity_performance_comparison}
\end{table}

\subsection{Network Setup}\label{network_setup}

\subsubsection*{UAV corridors and legacy ground users} We study a cellular network consisting of $19$ sites arranged in a hexagonal layout with inter-site distance (ISD) of $500$\,m. As per (\ref{entangled_BS_locations}) and (\ref{entangled_BS_orientations}), each site has three BSs located at the same positions with equally-separated azimuth orientations. The reference bearing for each site is set to $30^\circ$. Thus, the network consists of $57$ BSs where they all share the given height and maximum transmission power of $25$\,m and $43$\,dBm, respectively. 

There are two types of users being served by BSs: 
\begin{itemize}
    \item The ground users (GUEs) that are assumed to share the same height $h_G = 1.5$m and be spatially distributed over the region $Q_G = [-750, 750] \times [-750, 750]$ according to the density function $\lambda_G(\bm{q})$;
    \item The UAVs that are distributed over four predefined aerial routes/corridors $Q_U = Q_1 \cup Q_2 \cup Q_3 \cup Q_4$ according to the density function $\lambda_U(\bm{q})$, where $Q_1 = [-770, -730] \times [-1000, 1000] \times [135, 150]$, $Q_2 = [-1000, 1000] \times [-770, -730] \times [105, 120]$, $Q_3 = [-1000, 1000]\times[730, 770]\times[105, 120]$, and $Q_4 = [730, 770]\times [-1000, 1000] \times [135, 150]$, i.e., UAVs traverse in corridors $Q_1$ and $Q_4$ at altitudes between $135$\,m to $150$\,m while UAVs in corridors $Q_2$ and $Q_3$ fly at altitudes between $105$\,m to $120$\,m.
\end{itemize}
The mixture density function $\lambda(\bm{q})$ over the region $Q = Q_G \cup Q_U$ is given by $\lambda(\bm{q}) = r\lambda_G(\bm{q}) + (1 - r) \lambda_U(\bm{q})$ where $r \in [0, 1]$ makes a trade-off between prioritizing the optimization for ground users or UAVs. We consider a uniform distribution for $\lambda_U(\bm{q})$ and study two choices for $\lambda_G(\bm{q})$: A uniform distribution and a Gaussian mixture defined as $\sum_{i = 1}^{4} \pi_i \mathcal{N}(\bm{\mu}_i, \bm{\Sigma}_i)$ where:
\begin{align}\label{GMM_PDF_Definition}
    \!\pi_1 &= 0.35, \quad \!\bm{\mu}_1 = [-375, -225]^T, \quad \bm{\Sigma}_1 = 5\times 10^4 \bm{I}_2, \\ \label{GMM_PDF_Definition_1}
    \!\pi_2 &= 0.25, \quad \!\bm{\mu}_2 = [150, 375]^T, \quad \textrm{ } \textrm{ } \bm{\Sigma}_2 = 4.2\times 10^4 \bm{I}_2, \\ \label{GMM_PDF_Definition_2}
    \!\pi_3 &= 0.25, \quad \!\bm{\mu}_3 = [ 375, -375]^T, \quad \bm{\Sigma}_3 = 3.2\times 10^4 \bm{I}_2, \\ \label{GMM_PDF_Definition_3}
    \!\pi_4 &= 0.15, \quad \!\bm{\mu}_4 = [-300,  300]^T, \quad \bm{\Sigma}_4 = 3.8\times 10^4 \bm{I}_2. 
\end{align}

\begin{figure}[!t]
\centering
\subfloat[Uniform distribution of ground users.]{\includegraphics[width=\columnwidth]{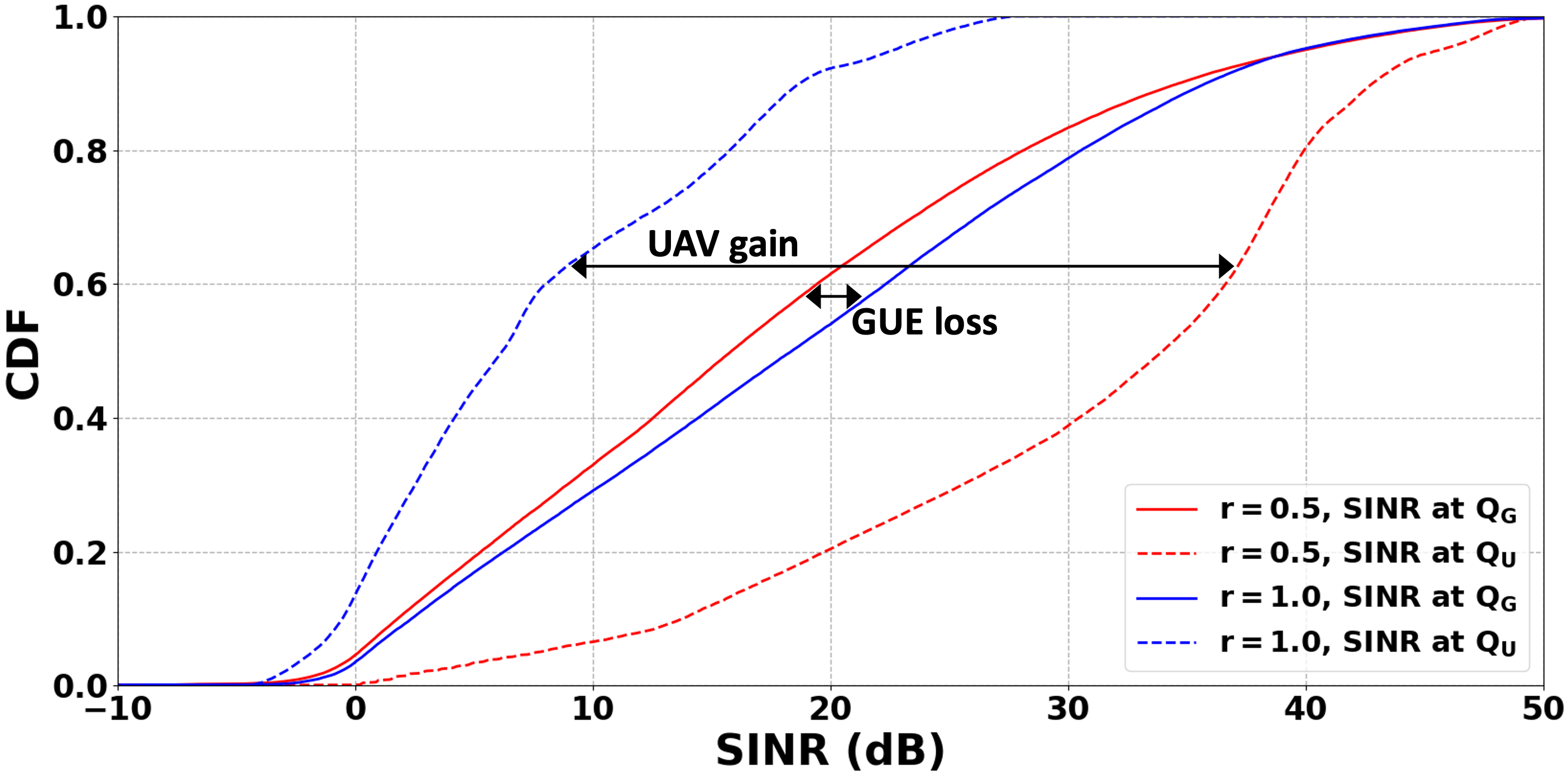}}
\hspace{0mm}\\
\vspace*{3mm}
\subfloat[Gaussian mixture distribution of ground users.]{\includegraphics[width=\columnwidth]{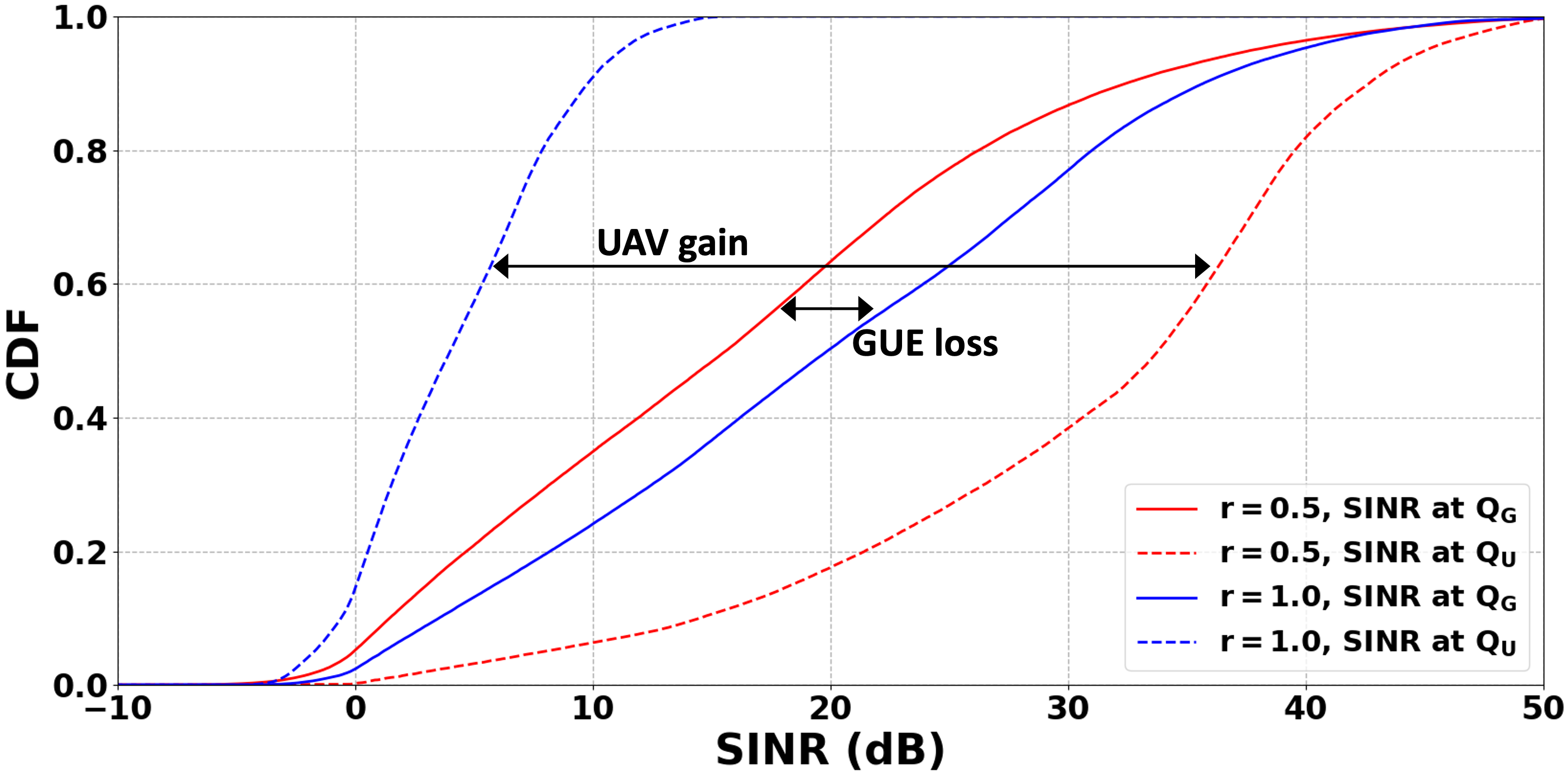}
}
\caption{The CDF of SINR when the cell partitioning, antenna tilts, transmission powers, and the deployment of new BSs are jointly optimized via Algorithm \ref{gamma_1_2_algorithm}. }
\label{sum_log_rate_CDF_algorithm2}
\end{figure}

\subsubsection*{Channel model} Following the 3GPP specifications \cite{3GPP38901,3GPP36777}, the parameters $a_{\bm{q}}$ and $b_{\bm{q}}$ are given for a carrier frequency of $2$\,GHz as follows:
\begin{equation}\label{a_q_values}
a_{\bm{q}} =
\begin{cases}
34.02\,\textrm{dB}, & \text{if}\ \bm{q}\in Q_U, \\
38.42\,\textrm{dB}, & \text{if}\ \bm{q}\in Q_G,
\end{cases}
\end{equation}
\begin{equation}\label{b_q_values}
b_{\bm{q}} =
\begin{cases}
22 \textrm{ (for a pathloss exponent of 2.2)}, & \text{if}\ \bm{q}\in Q_U, \\
30 \textrm{ (for a pathloss exponent of 3.0)}, & \text{if}\ \bm{q}\in Q_G.
\end{cases}
\end{equation}
Furthermore, the directional antennas have a maximum antenna gain of $A_{\max} = 14$\,dBi at the boresight and their vertical and horizontal half-power beamwidths are $\theta_{\textrm{3dB}} = 10^\circ$ and $\phi_{\textrm{3dB}} = 65^\circ$, respectively. Finally, the KPI-specific parameters $T$, $\beta$ and $o_n$ are set to $-5$, $0.5$ and $0.002$, respectively, for all $n \in \{1, \cdots, N\}$.

\begin{figure}[!t]
\centering
\subfloat[Optimal deployment and bearing of BSs overlaid on the heatmap representing the GUEs distribution.  
{Red stars indicate existing cellular sites (fixed location and bearings), blue circles indicate newly deployed sites (optimized location and bearings).} Numbers in the figure represent the indices of the $19$ sites. ]{\includegraphics[width=\columnwidth]{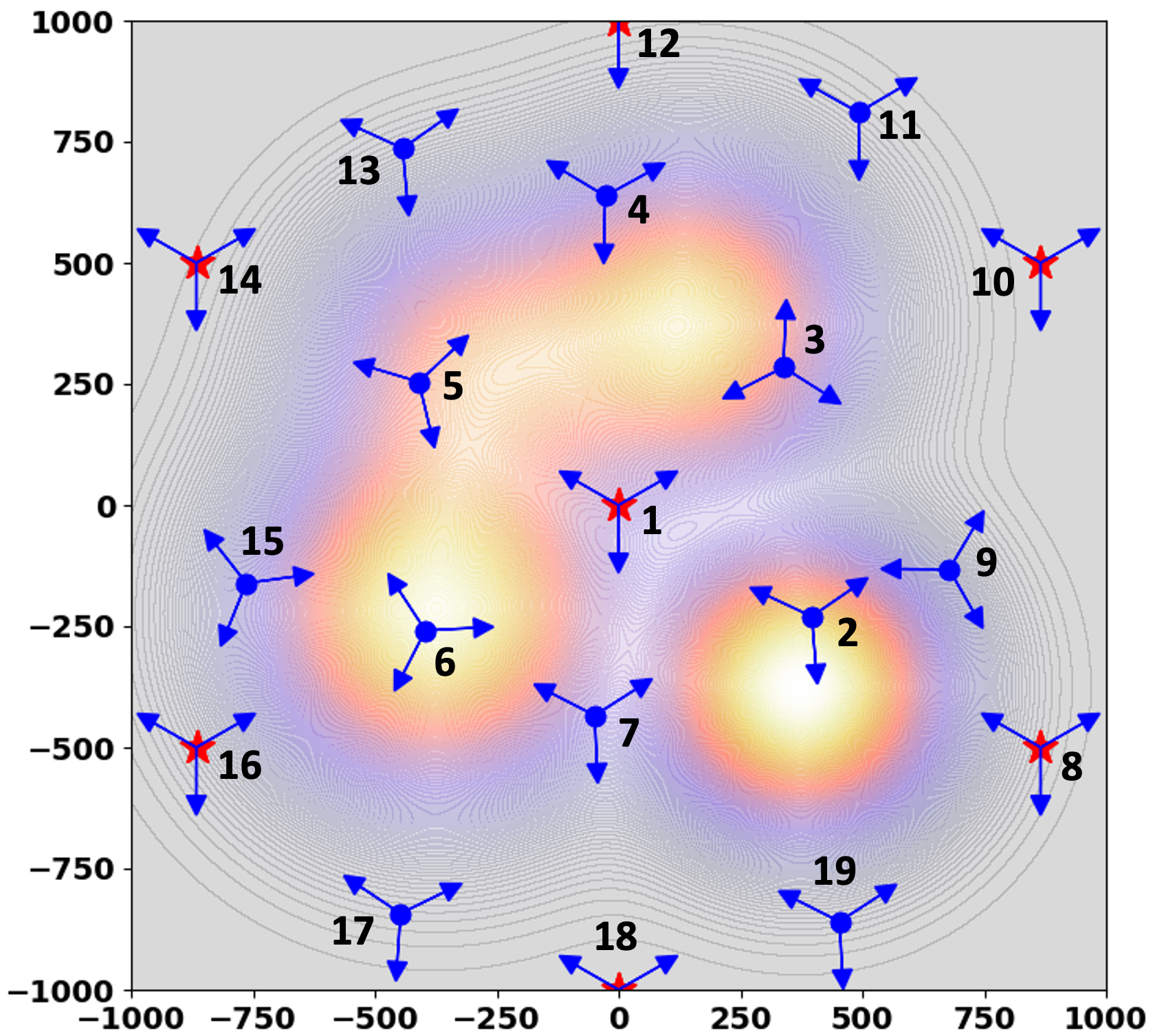}}
\hspace{0mm}\\
\vspace*{3mm}
\subfloat[Optimal cell partitioning for GUEs and UAV corridors.]{\includegraphics[width=\columnwidth]{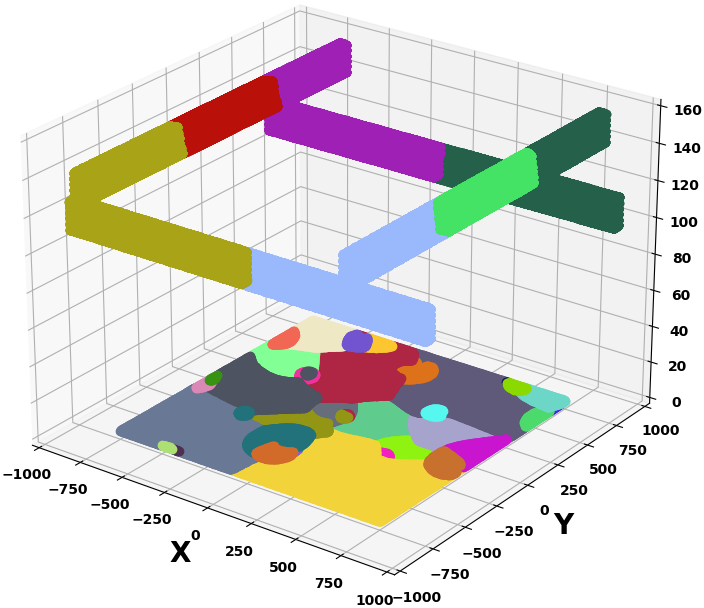}
}
\caption{Optimal deployment and cell partitioning when network is optimized via Algorithm \ref{gamma_1_2_algorithm} for $r = 0.5$ and GUEs are distributed according to the Gaussian mixture. }
\label{sum_log_rate_cell_partitioning_heatmap_azimuth_overlay}
\end{figure}

\subsection{Coverage-Capacity Optimization for GUEs and UAVs:}\label{Coverage_Capacity_Optimization_for_GUEs_and_UAVs}

\subsubsection*{Achieved KPI}\label{sum_log_rate_kpi_values}
Table \ref{coverage_capacity_performance_comparison} reports the final performance achieved by our proposed algorithm for the KPI \#1 defined in Section \ref{section:KPIs}. For both choices of uniform and Gaussian mixture distributions for ground users, we report the performance achieved (i) when the antenna tilts and power allocation of all existing BS are jointly optimized, as detailed in Algorithm \ref{gamma_1_1_algorithm}; (ii) when additionally a subset of BSs are optimally deployed and configured, as detailed in Algorithm \ref{gamma_1_2_algorithm}.  For the latter, we assume that all sites $m \in \mathcal{S} = \{1, 8, 10, 12, 14, 16, 18 \}$ are part of the existing infrastructure and fixed while the locations of all other sites $m \notin \mathcal{S}$ are optimized.  Table \ref{coverage_capacity_performance_comparison} shows that, regardless of the choice of GUE distribution, Algorithm \ref{gamma_1_2_algorithm} improves upon the performance achieved by Algorithm \ref{gamma_1_1_algorithm} since it further optimizes the location and bearing of new BSs.



\begin{figure}
\centering
\includegraphics[width=\columnwidth]{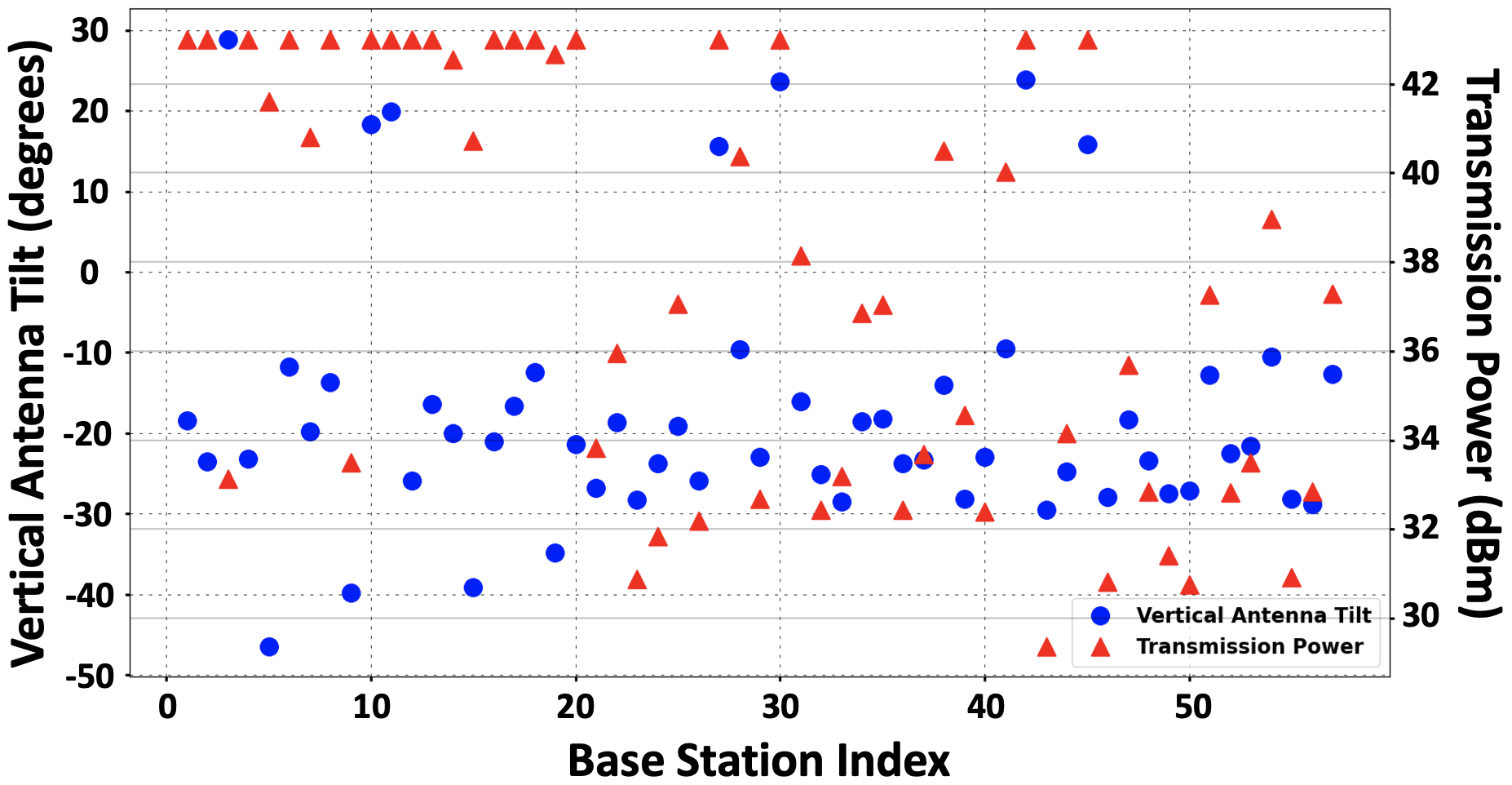}
\caption{Optimal BS antenna tilts (blue circles) and transmit powers (red triangles) for $r = 0.5$ when network is optimized via Algorithm \ref{gamma_1_2_algorithm}  and GUEs have Gaussian mixture distribution. }
\label{sum_log_rate_optimal_theta_powers}
\end{figure}



\subsubsection*{Resulting performance}\label{simulations_sum_log_rate_resulting_performance} 
Fig.~\ref{sum_log_rate_CDF_algorithm2} depicts the cumulative distribution function (CDF) of SINR perceived at network users when the cell partitioning, vertical antenna tilts, and the transmission power of all BSs in addition to locations and bearing of new BSs are jointly optimized via Algorithm \ref{gamma_1_2_algorithm}. Fig.~\ref{sum_log_rate_CDF_algorithm2}a shows the plots when GUEs are uniformly distributed while Fig.~\ref{sum_log_rate_CDF_algorithm2}b corresponds to the scenario in which GUEs follow the Gaussian mixture distribution defined in (\ref{GMM_PDF_Definition})-(\ref{GMM_PDF_Definition_3}). These plots demonstrate the benefits of optimizing the network for both GUEs and UAVs ($r = 0.5$) as opposed to GUEs only ($r = 1$) which is commonly done in traditional cellular networks. More specifically, as annotated in Figs. \ref{sum_log_rate_CDF_algorithm2}a and \ref{sum_log_rate_CDF_algorithm2}b, the gain in the SINR performance for UAVs (between the blue dash-dash and red dash-dash curves) is much larger than the loss in the SINR performance for GUEs (between the blue solid and red solid curves).  Our mathematical framework offers flexibility on the choice of $r$, which controls the prioritization of optimization between two user types. Figs. \ref{sum_log_rate_CDF_algorithm2}a and \ref{sum_log_rate_CDF_algorithm2}b demonstrate that the choice of $r = 0.5$ leads to 3D connectivity in the network for both user types and enables the network to provide coverage for UAVs without severe degradation in the signal quality at GUEs.


Next, we report the optimal BS deployment and configuration as well as the resulting cell partitioning for the scenario in which the location and bearing of new BSs are optimized via Algorithm \ref{gamma_1_2_algorithm} and GUEs are distributed according to the Gaussian mixture defined in (\ref{GMM_PDF_Definition})-(\ref{GMM_PDF_Definition_3}). 
Fig. \ref{sum_log_rate_cell_partitioning_heatmap_azimuth_overlay}a shows site indices and the locations and orientations of all BSs overlaid on a heatmap of the defined GUE density. Fig. \ref{sum_log_rate_cell_partitioning_heatmap_azimuth_overlay}b shows the resulting cell partitioning for GUEs and 3D UAV corridors. The corresponding optimal vertical antenna tilts and transmission powers are shown in  Fig. \ref{sum_log_rate_optimal_theta_powers}.

\begin{figure}[!t]
\centering
\subfloat[Uniform distribution of ground users.]{\includegraphics[width=\columnwidth]{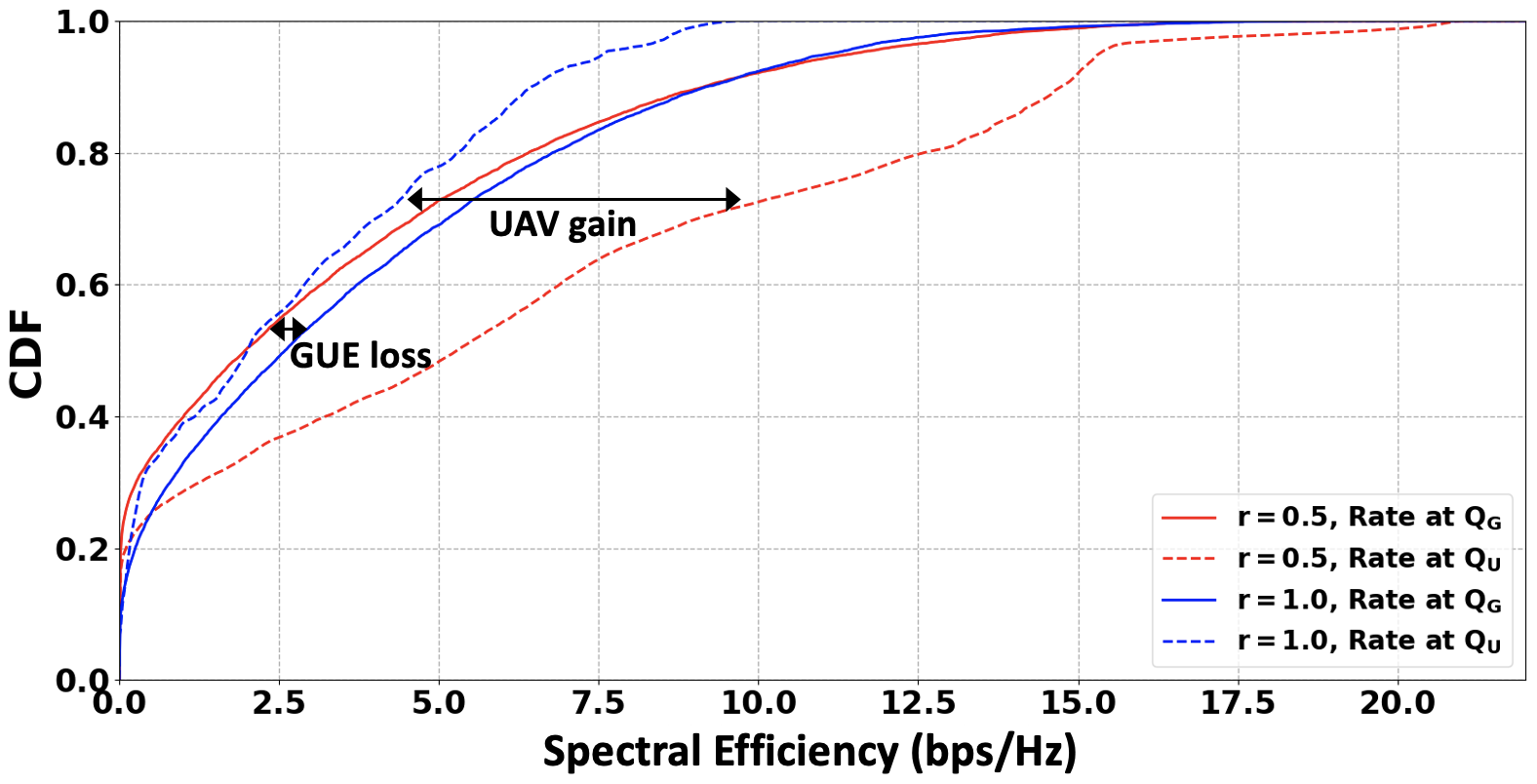}}
\hspace{0mm}\\
\vspace*{3mm}
\subfloat[Gaussian mixture distribution of ground users.]{\includegraphics[width=\columnwidth]{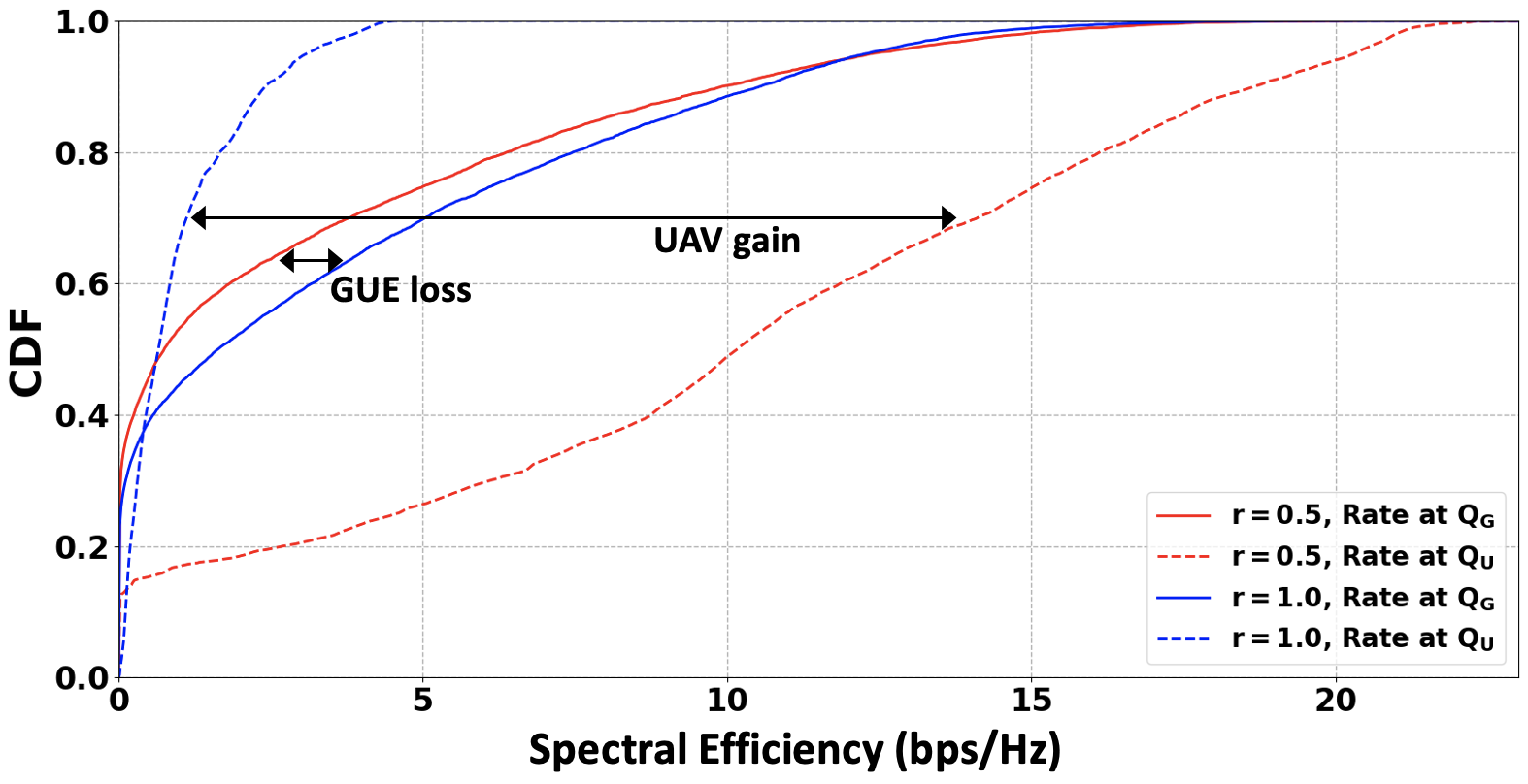}
}
\caption{The CDF of spectral efficiency (bps/Hz) when cell partitioning, antenna tilts, transmit powers, and the deployment of new BSs are jointly optimized via Algorithm \ref{gamma_2_2_algorithm}. }
\label{capacity_per_region_CDF_algorithm2}
\end{figure}

\subsection{Capacity-per-Region Optimization for GUEs and UAVs:}\label{Capacity_per_Region_Optimization_for_GUEs_and_UAVs}

\begin{table}[t!]
\centering
\caption{ Coverage-capacity performance comparison with and without the deployment of new BSs. Values are reported for $r = 0.5$ and two choices of distributions for GUEs. }
\begin{tabular}{|cc||ccc|}
 \toprule
\backslashbox{Algorithm}{GUE Distribution}         &\!\!\!\!\!\!&  Uniform  &\!\!\!\!\!\!&  Gaussian Mixture  \\\hline \hline
{\bf {\small Algorithm \ref{gamma_2_1_algorithm}}} &\!\!\!\!\!\!& $178.8899$&\!\!\!\!\!\!&    $176.3338$      \\ 
{\bf {\small Algorithm \ref{gamma_2_2_algorithm}}} &\!\!\!\!\!\!& $184.0236$&\!\!\!\!\!\!&    $190.2231$      \\ 
 \bottomrule 
\end{tabular}
  \label{capacity_per_region_performance_comparison}
\end{table}

\subsubsection*{Achieved KPI} 
For both choices of uniform and Gaussian mixture distributions for GUEs, we utilize (i) Algorithm \ref{gamma_2_1_algorithm} to find the optimal antenna tilts and transmission powers that maximizes KPI \#2 defined in Section \ref{section:KPIs}; (ii) Algorithm \ref{gamma_2_2_algorithm} to additionally optimize the locations and bearings of new BSs. 
The final performance values are summarized in Table \ref{capacity_per_region_performance_comparison}. Similar to the observation in Section \ref{sum_log_rate_kpi_values}, regardless of the choice of distribution for GUEs, Algorithm \ref{gamma_2_2_algorithm} improves the performance of Algorithm \ref{gamma_2_1_algorithm} by jointly optimizing the locations and bearing of new BSs.

\subsubsection*{Resulting performance}
The CDFs of spectral efficiency (bps/Hz) perceived at network users are shown in Figs. \ref{capacity_per_region_CDF_algorithm2}a and \ref{capacity_per_region_CDF_algorithm2}b when network parameters are optimized according to Algorithm \ref{gamma_2_2_algorithm} for uniform and Gaussian mixture distributions of GUEs, respectively. Similar to the observation made in Section \ref{simulations_sum_log_rate_resulting_performance}, regardless of the choice of distribution for GUEs, optimizing the network for both GUEs and UAVs ($r = 0.5$) leads to a significant gain in the rate performance for UAVs without causing a severe loss to the rate performance for GUEs. More specifically, by jointly optimizing the network for both GUEs and UAVs ($r = 0.5$) as opposed to GUEs only ($r = 1$), which is a common practice in traditional cellular networks, we can achieve 3D coverage and significantly improve the rate perceived at UAVs (between blue and red dash-dash curves) at the cost of small drop in the rate performance at GUEs (between blue and red solid curves).

\section{Conclusion}\label{Conclusion}

In this paper, we have developed a general mathematical framework for optimizing cell deployment and antenna configuration in wireless networks, leveraging the principles of quantization theory. Our framework extends the capabilities of traditional approaches by effectively addressing structured networks with deterministically located nodes. This enables precise modeling and optimization of network performance under controlled deployment scenarios.

To illustrate the capabilities of our framework, we provided two example applications. In the first one, we pursued the joint fine-tuning of antenna parameters for all BSs for cell shaping, i.e., to achieve optimal network coverage and capacity while ensuring efficient load balancing across cells. In the second one, we also tackled the strategic deployment of new BSs, optimizing both their locations and antenna parameters. We conducted these optimizations for a heterogeneous 3D population of network users, including legacy ground users and UAVs along aerial corridors.  Beyond these example applications, our framework accommodates the optimization of various system parameters for multiple key performance indicators over any deterministic (and potentially dynamic) 3D region of interest. 

To demonstrate its effectiveness, we presented case studies where we optimized two key metrics: the coverage-capacity trade-off and the capacity per region. These metrics were maximized for a heterogeneous user population, comprising ground users (GUEs) distributed either uniformly or following a Gaussian mixture, as well as UAVs along 3D aerial corridors.  The results of the case studies revealed that optimizing the locations and bearings of additional BS consistently outperformed approaches that solely focused on adjusting vertical antenna tilts and transmission powers, irrespective of the GUE distribution. Moreover, jointly optimizing the network for both GUEs and UAVs proved highly beneficial, significantly improving service for UAVs without causing severe performance degradation for GUEs.

\appendices


\section{Proof of Proposition \ref{theta_gradient_sumlograte_outage_function}}\label{Appendix_B}

The partial derivative $\frac{\partial}{\partial \theta_n} \mathcal{P}^{(1)}_{\gamma_1}$ is comprised of two terms: (1) the derivative of the integrand; and (2) the integral over the boundaries of $V_n$ and its neighboring regions. For any point $\bm{q}$ on the boundary of neighboring regions $V_n$ and $V_m$, the normal outward vectors have opposite directions but the same magnitude since  ${\sf RSS^{(n)}_{dBm}}(\bm{q}) = {\sf RSS^{(m)}_{dBm}}(\bm{q})$ according to Proposition \ref{optimal_cell_partitioning_sum_log_rate}; thus, the sum of elements in the second component is zero \cite{GuoJaf2016}.  The first component evaluates to:
\begin{multline}\label{appendix_B_eq_1}
    \frac{\partial \mathcal{P}^{(1)}_{\gamma_1}}{\partial \theta_n}  = \sum_{m=1}^{N} \int_{V_m} \bigg[ \beta \frac{\partial \gamma'^{(m)}(\bm{q})}{\partial \theta_n} \\ + (1 - \beta) \frac{\partial \widetilde{\gamma}''^{(m)}(\bm{q})}{\partial \theta_n} \bigg ] \lambda(\bm{q})d\bm{q}.
\end{multline}
The first term in (\ref{appendix_B_eq_1}) is given by:
\begin{align}\label{sum_log_rate_theta_gradient_eq}
&\frac{\partial \gamma'^{(m)}(\bm{q})}{\partial \theta_n} = 
\frac{\partial}{\partial \theta_n} \log_2\bigg( \log_2\Big(1 + {\sf SINR_{lin}^{(m)}}(\bm{q}) \Big) \bigg) \nonumber\\& = \frac{\log_2(e) \cdot \log_2(e) \cdot \ln{10} \cdot 0.1}{\log_2\big(1 + {\sf SINR_{lin}^{(m)}}(\bm{q})\big)}\times  \frac{ {\sf SINR_{lin}^{(m)}}(\bm{q})}{1 + {\sf SINR_{lin}^{(m)}}(\bm{q})} \nonumber\\& \times \frac{\partial}{\partial \theta_n} {\sf SINR_{dB}^{(m)}}(\bm{q}).
\end{align}
Additionally, we have:
\begin{align}\label{outage_theta_gradient_eq}
    \frac{\partial \widetilde{\gamma}''^{(m)}(\bm{q})}{\partial \theta_n} &=  \frac{\partial}{\partial \theta_n} \sigma\Big(\kappa \big[{\sf SINR_{dB}^{(m)}}(\bm{q}) - T \big] \Big) \nonumber \\ &
    = \kappa \cdot \sigma\Big(\kappa \big[{\sf SINR_{dB}^{(m)}}(\bm{q}) - T \big] \Big) \nonumber\\& \!\!\!\!\!\!\!\!\!\!\!\!\!\!\!\!\! \times  \bigg[1 - \sigma\Big(\kappa \big[{\sf SINR_{dB}^{(m)}}(\bm{q}) - T \big] \Big)\bigg]  \cdot  \frac{\partial {\sf SINR_{dB}^{(m)}}(\bm{q})}{\partial \theta_n},
\end{align}
where $\sigma(x) = \frac{1}{1 + e^{-x}}$ is the sigmoid function. Eq. (\ref{gamma_1_1_partial_derivative_wrt_theta}) in Proposition \ref{theta_gradient_sumlograte_outage_function} is then followed by substituting (\ref{sum_log_rate_theta_gradient_eq}) and (\ref{outage_theta_gradient_eq}) into (\ref{appendix_B_eq_1}). The expression in (\ref{partial_sinr_theta_derivatives}) for the term $\frac{\partial {\sf SINR_{dB}^{(m)}}(\bm{q})}{\partial \theta_n}$ follows from Proposition 5 in \cite{KarGerJaf2023}, concluding the proof.  $\hfill\blacksquare$

\section{Proof of Proposition \ref{power_gradient_sumlograte_outage_function}}\label{Appendix_C}

The partial derivative $\frac{\partial}{\partial \rho_n} \mathcal{P}^{(1)}_{\gamma_1}$ is comprised of two terms: (1) the derivative of the integrand; and (2) the integral over the boundaries of $V_n$ and its neighboring regions. It can be shown, using an argument similar to the one provided in Appendix \ref{Appendix_B}, that the second term amounts to zero.  Thus, the partial derivative is given by the following expression:
\begin{multline}\label{appendix_C_eq_1}
    \frac{\partial \mathcal{P}^{(1)}_{\gamma_1}}{\partial \rho_n}  = \sum_{m=1}^{N} \int_{V_m} \bigg[ \beta \frac{\partial \gamma'^{(m)}(\bm{q})}{\partial \rho_n} \\ + (1 - \beta) \frac{\partial \widetilde{\gamma}''^{(m)}(\bm{q})}{\partial \rho_n} \bigg ] \lambda(\bm{q})d\bm{q},
\end{multline}
in which
\begin{align}\label{sum_log_rate_power_gradient_eq}
&\frac{\partial \gamma'^{(m)}(\bm{q})}{\partial \rho_n} = \frac{\partial}{\partial \rho_n} \log_2\bigg( \log_2\Big(1 + {\sf SINR_{lin}^{(m)}}(\bm{q}) \Big) \bigg) \nonumber \\ & =
\frac{\log_2(e) \cdot \log_2(e) \cdot \ln{10} \cdot 0.1}{\log_2\big(1 + {\sf SINR_{lin}^{(m)}}(\bm{q})\big)}\times  \frac{{\sf SINR_{lin}^{(m)}}(\bm{q})}{1 + {\sf SINR_{lin}^{(m)}}(\bm{q})} \nonumber\\& \times \frac{\partial}{\partial \rho_n} {\sf SINR_{dB}^{(m)}}(\bm{q}),
\end{align}
and 
\begin{align}\label{outage_power_gradient_eq}
    \frac{\partial \widetilde{\gamma}''^{(m)}(\bm{q})}{\partial \rho_n} &=  \frac{\partial}{\partial \rho_n} \sigma\Big(\kappa \big[{\sf SINR_{dB}^{(m)}}(\bm{q}) - T \big] \Big) \nonumber \\ &
    = \kappa \cdot \sigma\Big(\kappa \big[{\sf SINR_{dB}^{(m)}}(\bm{q}) - T \big] \Big) \nonumber\\& \!\!\!\!\!\!\!\!\!\!\!\!\!\!\!\!\! \times  \bigg[1 - \sigma\Big(\kappa \big[{\sf SINR_{dB}^{(m)}}(\bm{q}) - T \big] \Big)\bigg]  \cdot  \frac{\partial {\sf SINR_{dB}^{(m)}}(\bm{q})}{\partial \rho_n},
\end{align}
where $\sigma(x) = \frac{1}{1 + e^{-x}}$ is the sigmoid function. Eq. (\ref{gamma_1_1_partial_derivative_wrt_power}) in Proposition \ref{power_gradient_sumlograte_outage_function} is then followed by substituting (\ref{sum_log_rate_power_gradient_eq}) and (\ref{outage_power_gradient_eq}) into (\ref{appendix_C_eq_1}). The expression in (\ref{partial_sinr_rho_derivatives}) for the term $\frac{\partial {\sf SINR_{dB}^{(m)}}(\bm{q})}{\partial \rho_n}$ follows from Proposition 6 in \cite{KarGerJaf2023}, concluding the proof.  $\hfill\blacksquare$

\section{Proof of Proposition \ref{convergence_gamma_1_1_algorithm}}\label{Appendix_D}

Proposition \ref{optimal_cell_partitioning_sum_log_rate} indicates that updating the cell $V_n$ according to (\ref{optimal_cell_partitioning_sum_log_rate_eq}), as it is done in  Algorithm \ref{gamma_1_1_algorithm}, yields the optimal cell partitioning for a given $\bm{\Theta}$ and $\bm{\rho}$; thus, the performance function $\mathcal{P}^{(1)}_{\gamma_1}$ will not decrease as a result of this update rule. Algorithm \ref{gamma_1_1_algorithm} updates the vertical antenna tilts $\bm{\Theta}$ and transmission powers $\bm{\rho}$ by following the gradient direction in small controlled steps, which does not decrease the performance function $\mathcal{P}^{(1)}_{\gamma_1}$. Hence, Algorithm \ref{gamma_1_1_algorithm} generates a sequence of non-decreasing performance function values. Since these values are also upper bounded (because of the limited transmission power available at each BS), the algorithm will converge. 
$\hfill\blacksquare$

\ifCLASSOPTIONcaptionsoff
  \newpage
\fi

\bibliographystyle{IEEEtran}
\bibliography{main}

\end{document}